\documentclass{pasj01}
\Received{$\langle$reception date$\rangle$}
\Accepted{$\langle$acception date$\rangle$}
\Published{$\langle$publication date$\rangle$}

\usepackage{color}

\begin{document}

\title{SILVERRUSH. III. Deep Optical and Near-Infrared \\
Spectroscopy for Ly$\alpha$ and UV-Nebular Lines of\\
Bright Ly$\alpha$ Emitters at $z=6-7$\\
}
\author{Takatoshi Shibuya\altaffilmark{1}, 
Masami Ouchi\altaffilmark{1,2}, 
Yuichi Harikane\altaffilmark{1,3}, 
Michael Rauch\altaffilmark{4}, 
Yoshiaki Ono\altaffilmark{1}, 
Shiro Mukae\altaffilmark{1,5}, 
Ryo Higuchi\altaffilmark{1,3}, 
Takashi Kojima\altaffilmark{1,3}, 
Suraphong Yuma\altaffilmark{6}, 
Chien-Hsiu Lee\altaffilmark{7}, 
Hisanori Furusawa\altaffilmark{8}, 
Akira Konno\altaffilmark{1,5}, 
Crystal L. Martin\altaffilmark{9}, 
Kazuhiro Shimasaku\altaffilmark{5,10}, 
Yoshiaki Taniguchi\altaffilmark{11}, 
Masakazu A. R. Kobayashi\altaffilmark{12}, 
Masaru Kajisawa\altaffilmark{13}, 
Tohru Nagao\altaffilmark{13}, 
Tomotsugu Goto\altaffilmark{14}, 
Nobunari Kashikawa\altaffilmark{8,15}, 
Yutaka Komiyama\altaffilmark{8,15}, 
Haruka Kusakabe\altaffilmark{5}, 
Rieko Momose\altaffilmark{14}, 
Kimihiko Nakajima\altaffilmark{16}, 
Masayuki Tanaka\altaffilmark{8,15}, 
and
Shiang-Yu Wang\altaffilmark{17}
}

\altaffiltext{1}{Institute for Cosmic Ray Research, The University of Tokyo, 5-1-5 Kashiwanoha, Kashiwa, Chiba 277-8582, Japan }
\altaffiltext{2}{Kavli Institute for the Physics and Mathematics of the Universe (Kavli IPMU, WPI), University of Tokyo, Kashiwa, Chiba 277-8583, Japan}
\altaffiltext{3}{Department of Physics, Graduate School of Science, The University of Tokyo, 7-3-1 Hongo, Bunkyo, Tokyo 113-0033, Japan} 
\altaffiltext{4}{Observatories of the Carnegie Institution of Washington, 813 Santa Barbara Street, Pasadena, CA 91101, USA}
\altaffiltext{5}{Department of Astronomy, Graduate School of Science, The University of Tokyo, 7-3-1 Hongo, Bunkyo, Tokyo 113-0033, Japan} 
\altaffiltext{6}{Department of Physics, Faculty of Science, Mahidol University, Bangkok 10400, Thailand }
\altaffiltext{7}{Subaru Telescope, NAOJ, 650 N Aohoku Pl., Hilo, HI 96720, USA}
\altaffiltext{8}{National Astronomical Observatory, Mitaka, Tokyo 181-8588, Japan}
\altaffiltext{9}{Department of Physics, University of California, Santa Barbara, CA, 93106, USA}
\altaffiltext{10}{Research Center for the Early Universe, Graduate School of Science, The University of Tokyo, 7-3-1 Hongo, Bunkyo, Tokyo 113-0033, Japan}
\altaffiltext{11}{The Open University of Japan, Wakaba 2-11, Mihama-ku, Chiba 261-8586, Japan}
\altaffiltext{12}{Faculty of Natural Sciences, National Institute of Technology, Kure College, 2-2-11 Agaminami, Kure, Hiroshima 737-8506, Japan}
\altaffiltext{13}{Research Center for Space and Cosmic Evolution, Ehime University, Bunkyo-cho 2-5, Matsuyama 790-8577, Japan}
\altaffiltext{14}{Institute of Astronomy, National Tsing Hua University, 101 Section 2, Kuang-Fu Road, Hsinchu 30013, Taiwan} 
\altaffiltext{15}{The Graduate University for Advanced Studies (SOKENDAI), 2-21-1 Osawa, Mitaka, Tokyo 181-8588}
\altaffiltext{16}{European Southern Observatory, Karl-Schwarzschild-Str. 2, D-85748, Garching bei Munchen, Germany}
\altaffiltext{17}{Academia Sinica, Institute of Astronomy and Astrophysics, 11F of AS/NTU Astronomy-Mathematics Building, No.1, Sec. 4, Roosevelt Rd, Taipei 10617, Taiwan}
\altaffiltext{\dag}{Based on data obtained with the Subaru Telescope. The Subaru Telescope is operated by the National Astronomical Observatory of Japan.}
\altaffiltext{\ddag}{This paper includes data gathered with the 6.5 meter Magellan Telescopes located at Las Campanas Observatory, Chile. }
\email{shibyatk@icrr.u-tokyo.ac.jp}


\maketitle

\begin{abstract}
We present Ly$\alpha$ and UV-nebular emission line properties of bright Ly$\alpha$ emitters (LAEs) at $z=6-7$ with a luminosity of $\log{L_{\rm Ly\alpha}/{\rm [erg\ s^{-1}]}} =43-44$ identified in the 21-deg$^2$ area of the SILVERRUSH early sample developed with the Subaru Hyper Suprime-Cam (HSC) survey data. Our optical spectroscopy newly confirm 21 bright LAEs with clear Ly$\alpha$ emission, and contribute to make a spectroscopic sample of 96 LAEs at $z=6-7$ in SILVERRUSH. From the spectroscopic sample, we select 7 remarkable LAEs as bright as Himiko and CR7 objects, and perform deep Keck/MOSFIRE and Subaru/nuMOIRCS near-infrared spectroscopy reaching the $3\sigma$-flux limit of $\sim 2\times 10^{-18}$ erg s$^{-1}$ for the UV-nebular emission lines of He {\sc ii}$\lambda1640$, C {\sc iv}$\lambda\lambda1548,1550$, and O {\sc iii}]$\lambda\lambda1661,1666$. Except for one tentative detection of C {\sc iv}, we find no strong UV-nebular lines down to the flux limit, placing the upper limits of the rest-frame equivalent widths ($EW_0$) of $\sim2-4$ \AA\ for C {\sc iv}, He {\sc ii}, and O {\sc iii}] lines. Here we also investigate the VLT/X-SHOOTER spectrum of CR7 whose $6 \sigma$ detection of He {\sc ii} is claimed by Sobral et al. Although two individuals and the ESO-archive service carefully re-analyze the X-SHOOTER data that are used in the study of Sobral et al., no He {\sc ii} signal of CR7 is detected, supportive of weak UV-nebular lines of the bright LAEs even for CR7. Spectral properties of these bright LAEs are thus clearly different from those of faint dropouts at $z\sim 7$ that have strong UV-nebular lines shown in the various studies. Comparing these bright LAEs and the faint dropouts, we find anti-correlations between the UV-nebular line $EW_0$ and UV-continuum luminosity, which are similar to those found at $z\sim 2-3$.
\end{abstract}

\section{Introduction}\label{sec_intro}

Bright Ly$\alpha$-emitting galaxies are important objects in the studies of the early Universe and galaxy formation. The bright Ly$\alpha$ emission with $\log{L_{{\rm Ly}\alpha}}/{\rm [erg\ s^{-1}]}\simeq43-44$ is expected to be reproduced in various physical mechanisms (e.g., \cite{2014Natur.505..186F,2015arXiv150607173P}). Very young and metal-free stars (Population III; Pop III) hosted in galaxies would emit the substantially strong Ly$\alpha$ radiation with a narrow He {\sc ii}$\lambda1640$ line ($\lesssim200$ km s$^{-1}$) and a Ly$\alpha$ equivalent width (EW) enhancement. On the other hand, active galactic nuclei (AGNs) would also produce the bright Ly$\alpha$ emission with high ionization metal lines such as N {\sc v}$\lambda\lambda1238, 1240$ and C {\sc iv}$\lambda\lambda1548, 1550$ due to the strong UV radiation from the central ionizing source. In addition, the highly-complex Ly$\alpha$ radiative transfer in the interstellar medium (ISM) makes it difficult to understand the Ly$\alpha$ emitting mechanism (e.g., \cite{1991ApJ...370L..85N,2006MNRAS.367..979H}). 

Ly$\alpha$ emitters (LAEs) have been surveyed by imaging observations with dedicated narrow-band (NB) filters. During the last decades, a wide FoV of Subaru/Suprime-Cam (SCam) has allowed us to identify LAE candidates at the bright-end of Ly$\alpha$ luminosity functions (LFs; e.g., \cite{2005PASJ...57..165T, 2006PASJ...58..313S, 2007ApJS..172..523M, 2008ApJ...677...12O, 2008ApJS..176..301O, 2010ApJ...723..869O, 2010ApJ...725..394H, 2006ApJ...648....7K, 2011ApJ...734..119K, 2014ApJ...797...16K, 2015MNRAS.451..400M}). Follow-up optical spectroscopic observations have confirmed several bright LAEs at $z\simeq6.6$ (e.g., {\it Himiko}: \cite{2009ApJ...696.1164O}; {\it CR7} and {\it MASOSA}: \cite{2015ApJ...808..139S}; {\it COLA1}: \cite{2016ApJ...825L...7H}; \cite{2017ApJ...837...11B}), and at $z\simeq5.7$ (\cite{2012ApJ...760..128M}). 

However, subsequent multi-band observations find the heterogeneity in the nature of these bright LAEs. \citet{2015MNRAS.451.2050Z} have reported no detections of He {\sc ii} nor C {\sc iv} from Himiko with VLT/X-SHOOTER. A deep ALMA observation reveals that Himiko has no strong $[$C {\sc ii}$]158\mu$m line and dust continuum emission \citep{2013ApJ...778..102O}. Combined with morphological properties, the bright Ly$\alpha$ emission of Himiko is probably caused by intense star formation in a galaxy merger. On the other hand,  \citet{2015ApJ...808..139S} have claimed that a narrow He {\sc ii} line is detected at the $6\sigma$ significance level from CR7 based on a deep VLT/X-SHOOTER near-infrared (NIR) spectroscopy. The He {\sc ii} detection might suggest that CR7 host Pop III stellar populations. Recently, a number of theoretical studies interpret the  strong He {\sc ii} emission from CR7 (e.g., \cite{2015MNRAS.453.2465P, 2016MNRAS.460.4003A, 2016MNRAS.462.2184H, 2016arXiv161100780J, 2016ApJ...823...74D, 2016ApJ...829L...6S, 2016MNRAS.460.3143S, 2016MNRAS.460L..59V, 2017MNRAS.468L..77P, 2017arXiv170100814V}). In contrast to the claim of the He {\sc ii} detection, CR7 clearly includes old stellar population found from analyses for photometric data (\cite{2016arXiv160900727B}), suggesting that this system would not be truly young. These studies indicate that the nature of bright LAEs has become a hot topic of debate. 

Even in the substantial observational and theoretical efforts, the diversity of the bright LAEs has not been unveiled yet due to the small statistics. In this paper, we present the results of our optical and NIR spectroscopic observations for bright LAEs selected with data of a new wide-FoV camera, Hyper Suprime-Cam (HSC), on the Subaru Telescope. In our spectroscopic observations, we newly identify 21 bright LAEs with $\log{L_{\rm Ly\alpha}}/{\rm [erg\ s^{-1}]}\simeq43-44$, which have enlarged the spectroscopic sample of bright LAEs by a factor of four. 

This is the third paper in our ongoing HSC research project for Ly$\alpha$-emitting objects, {\it Systematic Identification of LAEs for Visible Exploration and Reionization Research Using Subaru HSC} ({\it SILVERRUSH}). In this project, we study various properties of high-$z$ LAEs, e.g., LAE clustering (\cite{2017arXiv170407455O}), photometric properties of Ly$\alpha$ line EW and Ly$\alpha$ spatial extent (\cite{2017arXiv170408140S}), spectroscopic properties of bright LAEs (this study), Ly$\alpha$ LFs (\cite{2017arXiv170501222K}), and LAE overdensity (R. Higuchi et al. in preparation). This program is one of the twin programs. Another program is the study for dropouts, Great Optically Luminous Dropout Research Using Subaru HSC (GOLDRUSH), that is detailed in \citet{2017arXiv170406004O} and \citet{2017arXiv170406535H}. Source catalogs for the LAEs and dropouts will be presented on our project webpage at http://cos.icrr.u-tokyo.ac.jp/rush.html.

This paper has the following structure. In Section \ref{sec_targets}, we describe the HSC data and target selections of bright LAEs for our optical and NIR spectroscopy. Section \ref{sec_obs} presents details of the spectroscopic observations for the bright LAEs and the data reduction. In Section \ref{sec_results}, we investigate physical properties for bright LAEs at $z\simeq6$ using our statistical sample of bright LAEs. In Section \ref{sec_discuss}, we discuss the implications for galaxy formation and evolution. We summarize our findings in Section \ref{sec_summary}.

Throughout this paper, we adopt the concordance cosmology with $(\Omega_{\rm m}, \Omega_{\rm \Lambda}, h) = (0.3, 0.7, 0.7)$ (\cite{2016A&A...594A..13P}). All magnitudes are given in the AB system (\cite{1983ApJ...266..713O}).

\section{Targets for Spectroscopy}\label{sec_targets}

\subsection{Imaging Data}\label{sec_imaging}

In March 2014, the Subaru telescope has started a large-area NB survey with HSC in a Subaru strategic program (SSP; \cite{2017arXiv170405858A}). This survey will construct a sample of LAEs at $z\simeq 2.2, 5.7, 6.6$, and $7.3$ with four NB filters of ${\it NB}387$, ${\it NB}816$, ${\it NB}921$, and ${\it NB}101$. The statistical LAE sample allows us to study the LAE evolution and physical processes of the cosmic reionization. 

In this study, we use the HSC SSP S16A broadband (BB; \cite{kawanomoto2017}) and ${\it NB}921$ and ${\it NB}816$ data that are obtained in 2014-2016. Note that this HSC SSP S16A data is significantly larger than the first-released data in \citet{2017arXiv170208449A}. 

The HSC images were reduced with the HSC pipeline, {\tt hscPipe} 4.0.2 (\cite{bosch2017}) which is a code from the Large Synoptic Survey Telescope (LSST) software pipeline (\cite{2008arXiv0805.2366I, doi:10.1117/12.857297, 2015arXiv151207914J}). The photometric calibration is carried out with the PanSTARRS1 processing version 2 imaging survey data (\cite{2013ApJS..205...20M,2012ApJ...756..158S,2012ApJ...750...99T}). The details of the data reduction are provided in \citet{2017arXiv170208449A, bosch2017, 2017arXiv170208449A}. 

The ${\it NB}921$ (${\it NB}816$) filter has a central wavelength of $\lambda_c=9215$\AA\ ($8177$\AA) and an FWHM of 135\AA\ (113\AA), which traces the redshifted Ly$\alpha$ emission line at $z=6.580\pm0.056$ ($z=5.726\pm0.046$). The transmission curves and the detailed specifications of these NB filters are presented in \citet{2017arXiv170407455O}. The method of the transmission curve measurements is given by \citet{kawanomoto2017}. 

The HSC SSP S16A ${\it NB}921$ and ${\it NB}816$ data cover a total survey area of $\simeq21.2$ and $\simeq13.8$ deg$^2$, respectively. The survey area consists of two Ultradeep (UD) fields: UD-COSMOS, and UD-SXDS, and three Deep (D) fields: D-ELAIS-N1, D-DEEP2-3, and D-COSMOS. The FWHM of the typical seeing size is $\simeq0.\!\!^{\prime\prime}6$. The $5\sigma$ NB limiting magnitudes for the UD and D fields are typically $\simeq25.5$ and $\simeq25.0$ mag in a $1.\!\!^{\prime\prime}5$-diameter aperture, respectively. The details of the HSC NB data are presented in \citet{2017arXiv170408140S}. This HSC ${\it NB}921$ and ${\it NB}816$ data provide the largest NB survey area for $z\simeq5.7-6.6$ LAEs even before the completion of the SSP observation. 

\begin{figure*}[t!]
 \begin{center}
  \includegraphics[width=165mm]{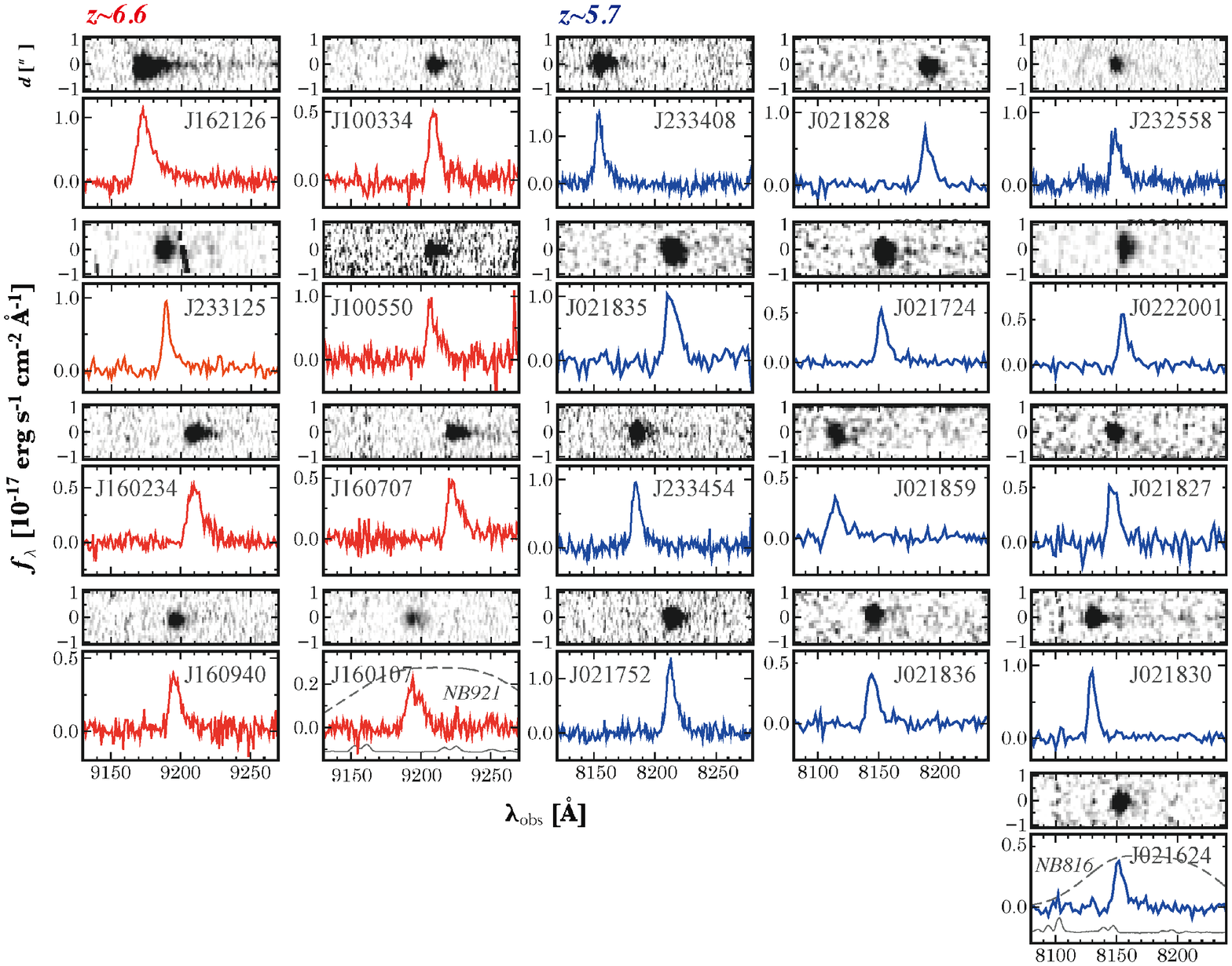}
 \end{center}
 \caption{Ly$\alpha$ spectra for the 21 newly-identified bright LAEs with $\log{L_{\rm Ly\alpha}}/{\rm [erg\ s^{-1}]} \simeq43-44$. The red and blue lines represent the Ly$\alpha$ spectra of the bright LAEs at $z\simeq6.6$ and $\simeq5.7$, respectively. The dashed gray curves indicate the transmission curves of ${\it NB}921$ and ${\it NB}816$. The solid gray lines denote the sky OH emission lines. The x-axis indicates the wavelength observed in air. The heliocentric motion of the Earth is not corrected in this figure.}\label{fig_hsc_spec_all}
\end{figure*}

\subsection{Selection of Bright LAEs}\label{sec_selection}

Using the HSC NB data, we select targets of bright LAE candidates with $\log{L_{{\rm Ly}\alpha}}/{\rm [erg\ s^{-1}]}\simeq43-44$ for follow-up spectroscopic observations. The details of the LAE selection are given in \citet{2017arXiv170408140S}, but we provide a brief description as follows. To identify objects with an NB magnitude excess in the HSC catalog, we apply the magnitude and color selection criteria similar to those of \citet{2008ApJS..176..301O} and \citet{2010ApJ...723..869O}. To remove spurious sources such as satellite trails and cosmic rays, we perform visual inspections to multi-band HSC images of $grizy$ and ${\it NB}$ for the objects selected in the magnitude and color selection criteria. We have also checked multi-epoch images to remove transients and asteroid-like moving objects. In total, photometric candidates of $1,153$ and $1,077$ LAEs at $z\simeq6.6$ and $z\simeq5.7$ are identified in the HSC-${\it NB}921$ and ${\it NB}816$ fields, respectively. Finally, we select bright LAE candidates with an NB magnitude of ${\it NB}\leq24$ mag corresponding to $\log{L_{{\rm Ly}\alpha}}/{\rm [erg\ s^{-1}]}\simeq43-44$.

\section{Spectroscopic Data}\label{sec_obs}

We carried out optical and NIR spectroscopic observations for the bright LAE candidates at $z\simeq5.7-6.6$ selected with the HSC NB data. These optical and NIR observations mainly 1) make spectroscopic confirmations through Ly$\alpha$ and 2) investigate properties of ionizing sources (e.g., the presence of metal-poor galaxies and AGN activity), respectively, for bright LAEs. Table \ref{tab_obs} summarizes the instruments, the exposure time and line flux limits of our spectroscopic observations for each target. 

In the following sections of Sections \ref{sec_obs_opt} and \ref{sec_obs_nir}, we describe the details of optical and NIR spectroscopic data, respectively.

\begin{figure*}[t!]
 \begin{center}
  \includegraphics[width=160mm]{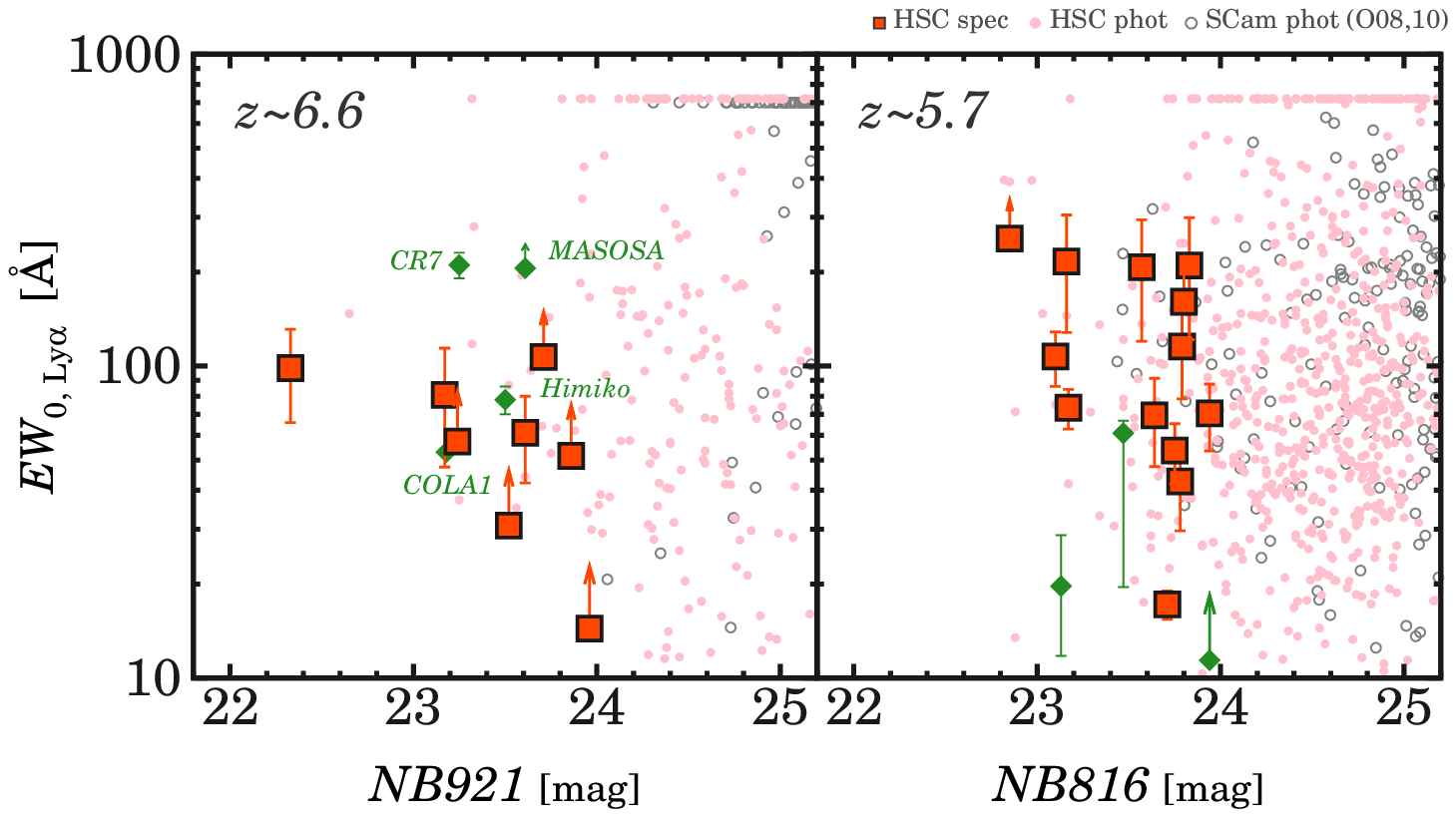}
 \end{center}
 \caption{NB magnitude and Ly$\alpha$ EW for LAEs at $z\simeq6.6$ (left) and $\simeq5.7$ (right). The red squares represent the 21 newly identified bright LAEs. The green diamonds indicate ${\it NB}<24$ bright LAEs which have been spectroscopically confirmed by previous studies (\cite{2009ApJ...696.1164O, 2012ApJ...760..128M, 2015ApJ...808..139S, 2016ApJ...825L...7H}). The magenta filled circles present LAE candidates in HSC LAE catalogs constructed by \citet{2017arXiv170408140S}. The gray open circles denote LAE candidates found in SCam NB surveys (\cite{2008ApJS..176..301O, 2010ApJ...723..869O}). The objects with $EW_{\rm 0,Ly\alpha}>700$\,\AA\ are plotted at $EW_{\rm 0,Ly\alpha}=700$\,\AA.}\label{fig_nb_lyaew_tile}
\end{figure*}

\subsection{Optical Spectroscopic Data}\label{sec_obs_opt}

We performed optical follow-up spectroscopy for bright LAE candidates at $z\simeq5.7-6.6$ to detect Ly$\alpha$ emission lines. The choice of the targets depends on the target visibility during the allocated time for individual spectroscopic observations. Basically, we selected the brightest LAE candidates as the targets in each observing run.

\subsubsection{Subaru/FOCAS}\label{sec_focas}

We used the Faint Object Camera and Spectrograph (FOCAS; \cite{2002PASJ...54..819K}) on the Subaru telescope to observe 16 LAE candidates. Out of the 16 objects, we observed 15 LAEs on 2016 June 21-22 and September 8 (S16A-060N and S16B-029N, PI: T. Shibuya), and one as a filler target of a FOCAS observation in 2015 December (S15B-059 in PI: S. Yuma; see \cite{2017arXiv170205107Y}). These observations were made with the VPH900 grism with the O58 order-cut filter, giving spectral coverage of $7500-10450$\,\AA\ with a dispersion of $0.74$\,\AA\ pix$^{-1}$. The $0.\!\!^{\prime\prime}8$-wide slit used gave a spectroscopic resolution of $R\simeq1800$ which is sufficient to distinguish $[$O {\sc ii}$]$ doublet lines from low-$z$ galaxy contaminants at $z\simeq0.6-0.8$. The observing nights were photometric, with good seeing of $\sim 0.\!\!^{\prime\prime}6 - 1.\!\!^{\prime\prime}0$. The Multi-Object Spectroscopy (MOS) mode was used to align securely the slits on our high-$z$ sources. Each of the $20$ minute exposures was taken by dithering the telescope pointing along the slit by $\pm 1.\!\!^{\prime\prime}0$. The standard star Feige34 was taken at the beginning and end of each observed night \citep{1990ApJ...358..344M}.

Our FOCAS spectra were reduced in a standard manner with the {\tt IRAF}\footnote{http:\/\/iraf.noao.edu\/} package (e.g., \cite{2006ApJ...648....7K, 2012ApJ...752..114S}). First, we performed flat-fielding with flat images, corrected for the image distortion, calibrated wavelengths with sky OH lines, and rejected sources illuminated by the cosmic ray injections. Next, we subtracted the sky background. After the sky background subtraction, we stacked the two-dimensional (2D) spectra. From each 2D data, we then extracted one-dimensional (1D) spectra using an extraction width of $\simeq \pm 0.\!\!^{\prime\prime}4-\pm 0.\!\!^{\prime\prime}8$ in the spatial direction of the slits. The extraction width is determined based on the extent of targets and the seeing conditions during the observations. Similarly, these extraction widths are used for the data obtained from the other optical and NIR spectrographs (Sections \ref{sec_obs_opt} and \ref{sec_obs_nir}). Finally, we carried out flux calibrations for the 1D spectra using the data of standard stars. 

The slit loss of the emission line flux has automatically been corrected in the flux calibration. This is because we observe the standard stars in the observing configuration (i.e. slit-width) and sky condition that are the same as those for our main targets. Note that our high-$z$ main targets are point source-like objects whose slit loss is the same as that of the standard stars. For this reason, we do not perform data reduction procedures for the slit loss correction for our optical and NIR spectra in Sections \ref{sec_obs_opt} and \ref{sec_obs_nir}.

\subsubsection{Magellan/LDSS3}\label{sec_ldss3}

We also used the Low Dispersion Survey Spectrograph 3 (LDSS3) on the Magellan II (Clay) telescope in October 2016 (PI: M. Rauch) to take spectroscopy for two bright LAE candidates. The seeing was $\sim 0.\!\!^{\prime\prime}6-1.\!\!^{\prime\prime}0$. We set the instrumental configuration to observe wavelength ranges of $8000-10000$\AA. The spatial pixel scale was $0.\!\!^{\prime\prime}189$ pix$^{-1}$, and the spectral dispersion was $0.47$\,\AA\ pix$^{-1}$. The slit width is $0.\!\!^{\prime\prime}8$.

\subsubsection{Magellan/IMACS}\label{sec_imacs}

In addition to the Subaru/FOCAS and Magellan/LDSS3 observations, we use spectroscopic data obtained with the Inamori-Magellan Areal Camera \& Spectrograph (IMACS; \cite{2011PASP..123..288D}) on the Magellan I Baade Telescope. The observations were conducted for high-$z$ galaxies in the SXDS field in 2007 - 2011 (PI: M. Ouchi; R. Higuchi et al in preparation). In the HSC LAE and IMACS catalog matching, we obtained optical spectra for eight bright LAEs.

\begin{figure*}[t!]
 \begin{center}
  \includegraphics[width=150mm]{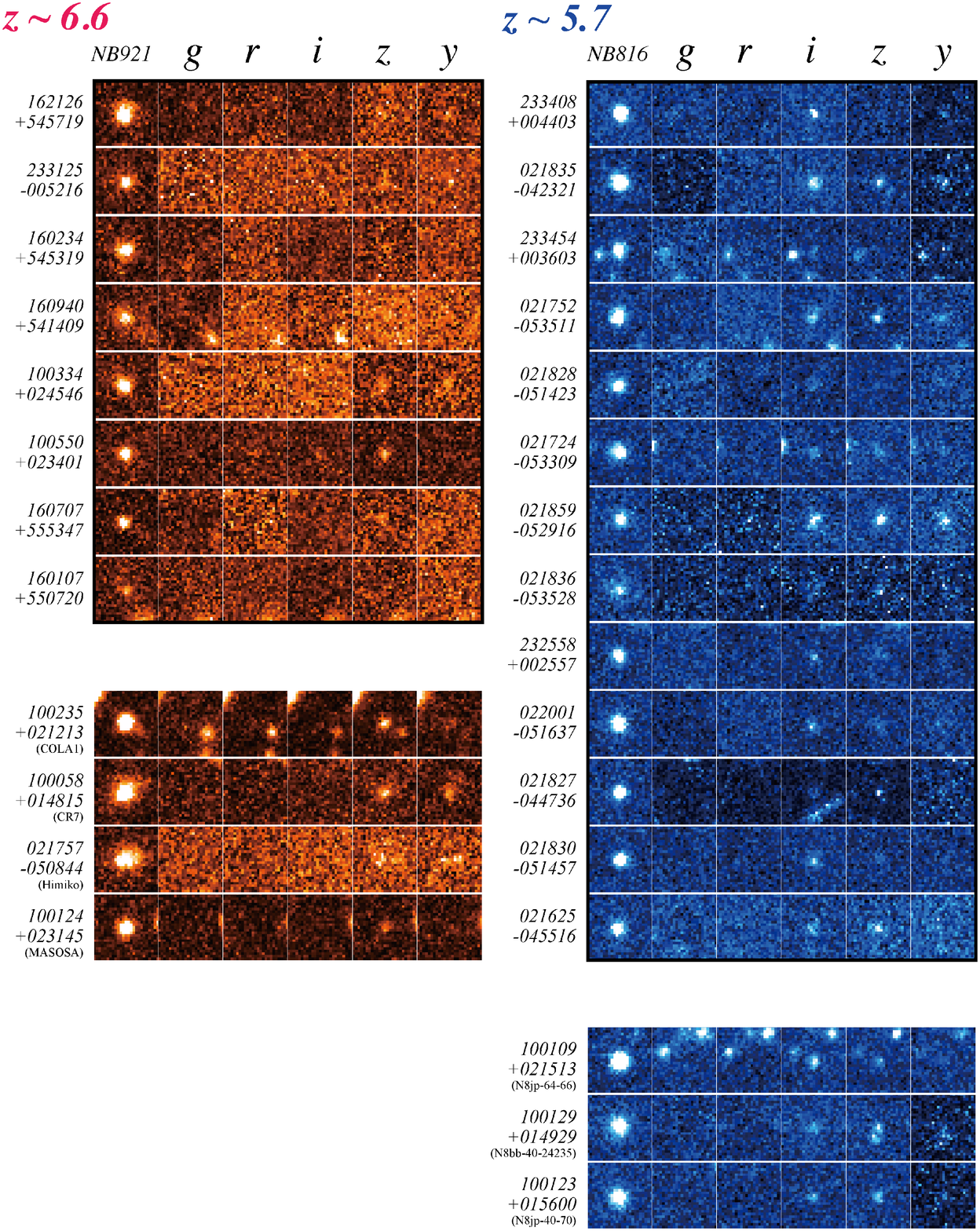}
 \end{center}
 \caption{HSC cutout images of the spectroscopically confirmed bright LAEs with ${\it NB}<24$ mag at $z\simeq6.6$ (left) and $\simeq5.7$ (right). Seven objects at the bottom are the previously identified bright LAEs at $z\simeq6.6$ (\cite{2016ApJ...825L...7H,2015ApJ...808..139S,2009ApJ...696.1164O}) and at $z\simeq5.7$ (\cite{2012ApJ...760..128M}). The image size is $4\arcsec \times4\arcsec$. The scale of flux density is arbitrary. }\label{fig_postage_all}
\end{figure*}

\subsubsection{LAE Spectroscopic Confirmations}\label{sec_confirm}

In total, we newly confirm 21 bright LAEs with a clear Ly$\alpha$ emission line in our Subaru/FOCAS and Magellan/LDSS3 observations and our Magellan/IMACS data. The 1D and 2D optical spectra of the 21 bright LAEs are shown in Figure \ref{fig_hsc_spec_all}. A prominent asymmetric emission line is found at $\simeq9210$\AA\, and $\simeq8160$\AA\, for each LAE at $z\simeq6.6$ and $z\simeq5.7$, respectively. These emission lines are detected at the $\simeq10-20\sigma$ significance levels. No other emission line feature is found in the range of observed wavelengths. We obtain the redshift of the bright LAEs by fitting the symmetric Gaussian profile to the observed Ly$\alpha$ emission lines in the wavelength ranges where the flux drops to $70$\% of its peak value (\cite{2014ApJ...788...74S}). Figure \ref{fig_nb_lyaew_tile} shows the NB magnitude and Ly$\alpha$ EW which is obtained in Section \ref{sec_phot}. As shown in Figure \ref{fig_nb_lyaew_tile}, our newly confirmed bright LAEs are as bright as e.g., Himiko and CR7. 

We also check whether our LAEs selected with the HSC data, HSC LAEs, are spectroscopically confirmed in previous studies for the COSMOS and SXDS fields (\cite{2007ApJS..172..523M, 2009ApJ...701..915T, 2008ApJS..176..301O,2010ApJ...723..869O, 2012ApJ...760..128M, 2015ApJ...808..139S, 2016ApJ...825L...7H}). In spectroscopic samples obtained by the previous studies, we find that 7 bright LAEs with ${\it NB}<24$ mag and 69 faint ones with ${\it NB}>24$ mag. In total, 96 LAEs are confirmed in our spectroscopic observations and the previous studies. Table \ref{tab_n_spec} summarizes the number of the spectroscopically confirmed HSC LAEs. 

The photometric properties and the HSC images for the bright LAEs in Table \ref{tab_phot} and Figure \ref{fig_postage_all}, respectively. Although most of the bright LAEs are not detected in blue bands of $g$ and $r$, COLA1 is marginally detected in the $r$-band image at $\simeq2.5\sigma$. 

Combining our 21 newly identified and the 7 previously confirmed bright LAEs (i.e., Himiko, CR7, MASOSA, COLA1, and three $z\simeq5.7$ \authorcite{2012ApJ...760..128M}'s objects), we construct a sample of 28 bright LAEs. The HSC data and our observations have enlarged a spectroscopic sample of bright LAEs by a factor of four. The large sample allows for a statistical study on physical properties of bright LAEs with $\log{L_{\rm Ly\alpha}}/{\rm [erg\ s^{-1}]}\simeq43-44$.

\subsubsection{Contamination Rates in the LAE Candidates}\label{sec_contami}

\begin{figure*}[t!]
 \begin{center}
  \includegraphics[width=130mm]{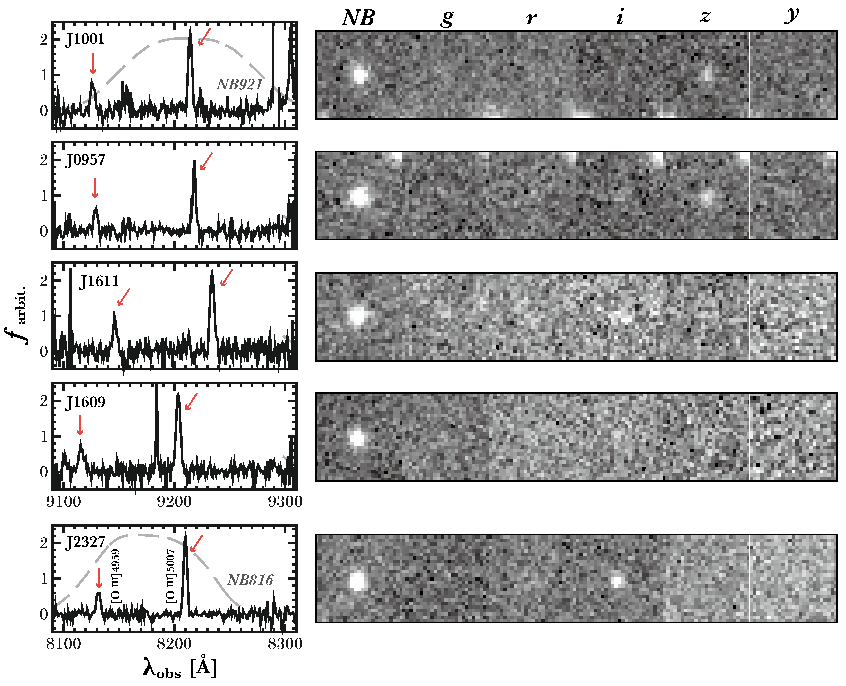}
 \end{center}
 \caption{(Left) Spectra for the 5 low-$z$ contaminants in the Subaru/FOCAS and Magellan/LDSS3 observations. The red arrows indicate the positions of [O {\sc iii}]$\lambda\lambda$4959, 5007 emission lines. The dashed gray curves denote the transmission curves of ${\it NB}921$ and ${\it NB}816$. The x-axis indicates the wavelength observed in air. The heliocentric motion of the Earth is not corrected in this figure. The y-axis represents the flux density in arbitrary units. (Right) HSC cutout images of the low-$z$ contaminants. The image size is $4^{\prime\prime}\times 4^{\prime\prime}$. The scale of flux density is arbitrary. The photometric properties of the low-$z$ contaminants are summarized in Table \ref{tab_lowz}.}\label{fig_lowz_all}
\end{figure*}

We estimate contamination rates, $f_{\rm contami}$, in the HSC LAE candidates using the spectroscopic data.  In our Subaru/FOCAS and Magellan/LDSS3 observations for 12 $z\simeq6.6$ and 6 $z\simeq5.7$ bright LAE candidates with ${\rm NB}<24$ mag, we identify 4 and 1 low-$z$ contaminants, respectively. Figure \ref{fig_lowz_all} presents the spectra and HSC cutout images for the low-$z$ contaminants. All of the 5 contaminants are strong [O {\sc iii}]$\lambda\lambda4959, 5007$ emitters at $z\simeq0.6-0.8$ with faint BB magnitudes. The H$\beta$ and H$\gamma$ emission lines are not significantly detected in the short integration time (i.e. $\simeq20-40$ minutes) of the FOCAS and LDSS3 observations. The photometric properties of these low-$z$ contaminants are listed in Table \ref{tab_lowz}. We find that $f_{\rm contami}\simeq33$\% ($=4/12$) and $\simeq17$\% ($=1/6$) for bright LAE candidates with ${\rm NB}<24$ mag at $z\simeq6.6$ and $z\simeq5.7$, respectively. 

We also calculate $f_{\rm contami}$ in the HSC LAE candidates including our Magellan/IMACS spectroscopic data (Section \ref{sec_imacs}). This spectroscopic sample includes faint HSC LAE candidates with ${\rm NB}>24$ mag. Combining our Subaru/FOCAS and Magellan/LDSS3 data and the cross-matching of the  Magellan/IMACS spectroscopic catalogs, we find that 28 and 53 HSC LAE candidates at $z\simeq6.6$ and $z\simeq5.7$ are spectroscopically observed. In total, we find that 4 out of 28 (4 out of 53) HSC LAE candidates are low-$z$ contaminants, and estimate $f_{\rm contami}$ to be $\simeq14$\% and $\simeq8$\% for the samples of $z\simeq6.6$ and $z\simeq5.7$ LAEs, respectively. 

In these estimates with the spectroscopic data, we find that $f_{\rm contami}\simeq0-30$\%. Table \ref{tab_contami} summarizes the contamination rates. These $f_{\rm contami}$ values are used for the contamination correction for e.g., LAE clustering (\cite{2017arXiv170407455O}), Ly$\alpha$ LFs (\cite{2017arXiv170501222K}), and LAE overdensity (R. Higuchi et al. in prep.).

\subsection{NIR Spectroscopic Data}\label{sec_obs_nir}

We performed deep NIR spectroscopy to investigate whether the rest-frame UV-nebular emission lines (i.e., C {\sc iv}$\lambda\lambda1548, 1550$, He {\sc ii}$\lambda1640$, and O {\sc iii}]$\lambda\lambda1661, 1666$) exist in bright LAEs. As a first attempt, we observed seven out of the spectroscopically confirmed 21 bright LAEs. The LAEs observed by NIR spectrographs are listed in Table \ref{tab_obs}. The choice of the targets depends on the target visibility during the allocated time for individual spectroscopic observations. Basically, we have selected the brightest LAEs as the targets in each observing run.

\subsubsection{Keck/MOSFIRE}\label{sec_mosfire}

We used the Multi-Object Spectrometer For Infra-red Exploration (MOSFIRE; \cite{2012SPIE.8446E..0JM}) on the Keck I telescope to observe 4 LAEs on 2016 September 9 (S16B-029N, PI: T. Shibuya) and an LAE on 2015 January 3-4 as a filler target (S15B-075, PI: M. Ouchi). 
Similar to the Subaru/FOCAS observations, the MOS mode was utilized to align securely the slits on our high-$z$ sources. We used the {\it Y} and {\it J} band filters for LAEs at $z\simeq5.7$ and $z\simeq6.6$, respectively. The seeing size was $\sim0.\!\!^{\prime\prime}5$-$0.\!\!^{\prime\prime}6$. The $0.\!\!^{\prime\prime}8$-wide slit was used, giving a spectral resolution of $R\simeq3500$. 

The data of objects and standard stars were reduced using the MOSFIRE data reduction pipeline.\footnote{https://keck-datareductionpipelines.github.io\/MosfireDRP\/} We conducted standard reduction processes for the MOSFIRE spectra with sets of default pipeline parameters (see e.g., \cite{2016arXiv160503436K}). Using $A$-spectral type stars which were taken in this observing run, we performed flux calibrations for the spectra of the target LAEs.

\renewcommand{\tabcolsep}{3pt}
\begin{longtable}{*{7}{c}}
\caption{Our Optical and NIR Spectroscopic Observations for Bright LAEs}\label{tab_obs}
\hline
Object ID & Opt. Inst. & $T_{\rm exp, opt}$ & $f_{\rm lim, opt}$ &  NIR. Inst. & $T_{\rm exp, NIR}$ & $f_{\rm lim, NIR}$  \\
 & & (minutes) & (erg s$^{-1}$ cm$^{-2}$) & & (minutes) & (erg s$^{-1}$ cm$^{-2}$)  \\
(1) & (2) & (3) & (4) & (5) & (6) & (7) \\
\hline
\endfirsthead
\endhead
\hline
\endfoot
\hline
\multicolumn{7}{l}{(1) Object ID. Sorted by the NB magnitude. } \\
\multicolumn{7}{l}{(2) Instrument for optical spectroscopy. } \\
\multicolumn{7}{l}{(3) Integration time for optical spectroscopy.} \\
\multicolumn{7}{l}{(4) The $1\sigma$ line flux sensitivity near Ly$\alpha$ emission lines.} \\
\multicolumn{7}{l}{(5) Instrument for NIR spectroscopy. } \\
\multicolumn{7}{l}{(6) Integration time for NIR spectroscopy.} \\
\multicolumn{7}{l}{(7) Average values of the $1\sigma$ line flux sensitivity at the expected wavelengths of C {\sc iv}, He {\sc ii}, and O {\sc iii}]. } \\
\multicolumn{7}{l}{$^a$ Spectroscopically confirmed with Magellan/IMACS. See Section \ref{sec_imacs}. } \\
\endlastfoot
HSC J162126$+$545719 & FOCAS & 60 & $\simeq1.2\times10^{-18}$  & MOSFIRE & 120 & $\simeq1.8\times10^{-18}$ \\ 
HSC J233125$-$005216 & LDSS3 & 90 & $\simeq0.5\times10^{-18}$ & --- & --- & --- \\
HSC J160234$+$545319 & FOCAS & 60 & $\simeq0.3\times10^{-18}$ & nuMOIRCS & 180 & $\simeq5.3\times10^{-18}$ \\ 
HSC J160940$+$541409 & FOCAS & 60 & $\simeq0.6\times10^{-18}$ & nuMOIRCS & 300 & $\simeq6.0\times10^{-18}$ \\ 
HSC J100334$+$024546 & FOCAS & 100 & $\simeq1.3\times10^{-18}$ & --- & --- & --- \\
HSC J100550$+$023401 & FOCAS & 60 & $\simeq1.0\times10^{-18}$ & MOSFIRE & 120 & $\simeq0.3\times10^{-18}$ \\ 
HSC J160707$+$555347 & FOCAS & 60 & $\simeq0.5\times10^{-18}$ & --- & --- & --- \\
HSC J160107$+$550720 & FOCAS & 60 & $\simeq0.3\times10^{-18}$ & --- & --- & --- \\\hline 
HSC J233408$+$004403 & FOCAS & 60 & $\simeq0.3\times10^{-18}$ & MOSFIRE & 120 & $\simeq0.8\times10^{-18}$ \\ 
HSC J021835$-$042321$^a$ & --- & --- & --- & MOSFIRE & 120 & $\simeq1.5\times10^{-18}$ \\ 
HSC J233454$+$003603 & FOCAS & 60 & $\simeq1.0\times10^{-18}$  & MOSFIRE & 120 & $\simeq0.6\times10^{-18}$ \\ 
HSC J021752$-$053511 & FOCAS & 60 & $\simeq0.1\times10^{-18}$  & --- & --- & --- \\
HSC J232558$+$002557 & FOCAS & 60 & $\simeq0.2\times10^{-18}$  & --- & --- & --- \\
HSC J022001$-$051637 & LDSS3 & 45 & $\simeq0.3\times10^{-18}$  & --- & --- & --- \\
\hline
\end{longtable}

\subsubsection{Subaru/nuMOIRCS}\label{sec_moircs}

We used the upgraded version of the Multi-Object InfraRed Camera and Spectrograph (nuMOIRCS; \cite{2006SPIE.6269E..16I,2008PASJ...60.1347S,2016SPIE.9908E..28F,2016SPIE.9908E..2GW}) on the Subaru telescope on 2016 June 21-22 to observe 2 LAEs at $z\simeq6.6$ (S16A-060N, PI: T. Shibuya). The MOS mode was used to align securely the slits on our high-$z$ sources. There were thin sky cirrus, but the weather condition was photometric. The seeing size was $\sim 0.\!\!^{\prime\prime}5$-$1.\!\!^{\prime\prime}0$. The width of each slit in the MOS masks is $0.\!\!^{\prime\prime}8$. We used the VPH-J grism, giving the spectral resolution of $R\simeq3000$. The standard star HIP115119 was observed on each night for flux calibrations. 

We reduced the nuMOIRCS spectra with {\tt IRAF} in the manner similar to the FOCAS data reduction (Section \ref{sec_focas}). We performed bias subtraction, flat fielding, image distortion correction, cosmic ray rejection, wavelength calibration, sky subtraction, and flux calibration.

\begin{figure}[t!]
 \begin{center}
  \includegraphics[width=80mm]{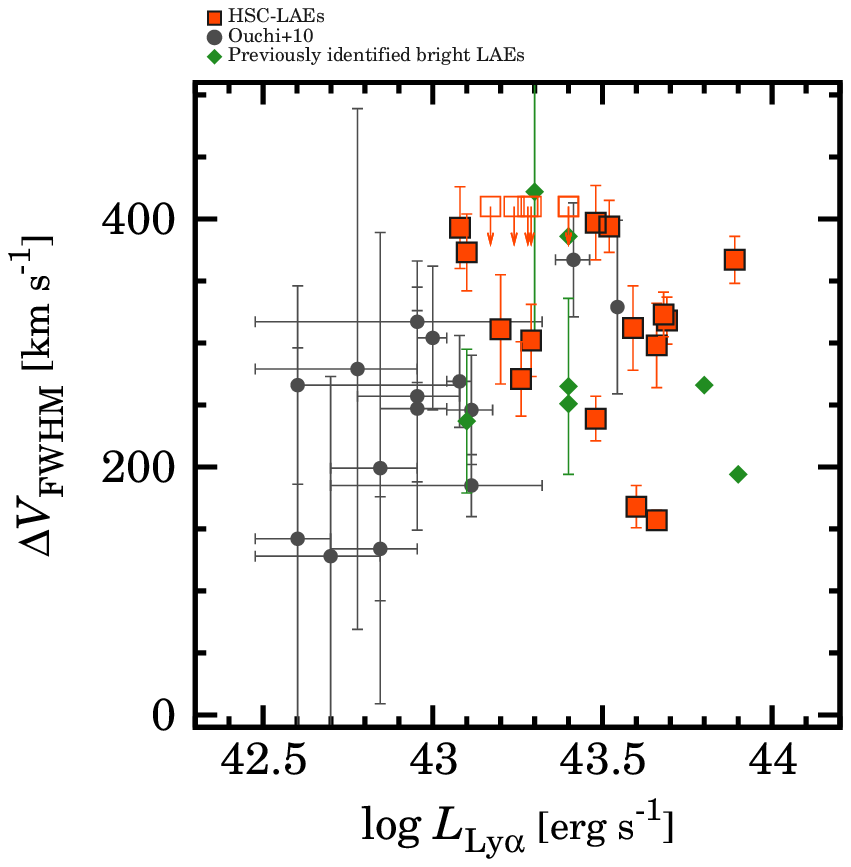}
 \end{center}
 \caption{Ly$\alpha$ line FWHM as a function of Ly$\alpha$ luminosity. The red filled and open squares indicate our bright LAEs whose Ly$\alpha$ emission line is spectroscopically resolved and not revolved, respectively. The green filled diamonds denote bright LAEs which have been previously confirmed (\cite{2016ApJ...825L...7H, 2015ApJ...808..139S, 2009ApJ...696.1164O, 2012ApJ...760..128M}). The black circles present faint $z\simeq6.6$ LAEs with $\log{L_{\rm Ly\alpha}}/{\rm [erg\ s^{-1}]}\lesssim43$ in \citet{2010ApJ...723..869O}. }\label{fig_phys_fwhm}
\end{figure}

\section{Results}\label{sec_results}

\subsection{Physical Properties}\label{sec_phot}

We present physical quantities related to the Ly$\alpha$ emission: Ly$\alpha$ flux, $f_{\rm Ly\alpha}$, Ly$\alpha$ luminosity, $L_{\rm Ly\alpha}$, and the rest-frame Ly$\alpha$ EW, $EW_{\rm 0, Ly\alpha}$, for the bright LAEs with a spectroscopic redshift. To obtain these quantities, we scale the observed Ly$\alpha$ spectra to match the NB and BB magnitudes. Here we assume the rest-frame UV spectral slope of $\beta=-2$. The $\beta$ parameter is defined by $f_\lambda\propto\lambda^\beta$ where $f_\lambda$ is a galaxy spectrum at $\simeq 1500-3000$\AA. The $2\sigma$ lower limits of $y$ ($z$)-band magnitudes are used for $z\simeq6.6$ ($z\simeq5.7$) LAEs whose UV continuum emission is not detected. For HSC J162126$+$545719 whose UV continuum is detected in the spectroscopic data (see Figure \ref{fig_hsc_spec_all}), we use the UV continuum flux density in the spectra to measure the $EW_{\rm 0, Ly\alpha}$ and $M_{\rm UV}$ values. Table \ref{tab_phys} presents the quantities of $f_{\rm Ly\alpha}$, $L_{\rm Ly\alpha}$, and $EW_{\rm 0, Ly\alpha}$ for our 21 bright LAEs including a sample of 7 LAEs identified by previous studies (\cite{2009ApJ...696.1164O, 2012ApJ...760..128M, 2015ApJ...808..139S, 2016ApJ...825L...7H}). Figure \ref{fig_nb_lyaew_tile} shows $EW_{\rm 0, Ly\alpha}$ as a function of NB magnitude. The $EW_{\rm 0, Ly\alpha}$ value ranges from $\simeq10$\,\AA\ to $\simeq300$\,\AA. 

Table \ref{tab_phys} also shows whether the bright LAEs are spatially extended or not in Ly$\alpha$ based on our measurements of isophotal areas, $A_{\rm iso}$ (see \cite{2017arXiv170408140S}). We find that only 5 out of the 28 bright LAEs show the spatially extended Ly$\alpha$ emission. The $A_{\rm iso}$ measurements indicate that Ly$\alpha$ emission of the bright LAEs is typically compact.

\begin{table}
  \tbl{Number of Spectroscopically Confirmed HSC LAEs at $z\simeq5.7-6.6$.}{%
  \begin{tabular}{ccc}
    \hline
Sample & $N_{\rm LAE,spec}$ & Spec. Obs. or Sample \\
(1) & (2) & (3) \\ \hline
Bright (${\it NB}<24$) & 13 & FOCAS, LDSS3 \\ 
Bright (${\it NB}<24$) &  8 & IMACS \\
Bright (${\it NB}<24$) &  7 & Literature$^a$ \\ 
Faint (${\it NB}>24$)$^b$ & 68 & LDSS3, IMACS, Literature$^a$ \\ \hline
Total & 96 & --- \\ \hline
  \end{tabular}}\label{tab_n_spec}
  \begin{tabnote}
    (1) LAE sample. (2) Number of spectroscopically confirmed LAEs. (3) Instruments for observations and spectroscopic samples. \\
    $^a$ \citet{2007ApJS..172..523M, 2009ApJ...701..915T, 2008ApJS..176..301O,2010ApJ...723..869O, 2012ApJ...760..128M, 2015ApJ...808..139S, 2016ApJ...825L...7H}. \\
    $^b$ See Tables \ref{tab_faint_z66} and \ref{tab_faint_z57}. 
  \end{tabnote}
\end{table}

\begin{table}
  \tbl{Contamination Rates in the HSC LAE Candidates}{%
  \begin{tabular}{ccccc}
    \hline
Redshift & $N_{\rm obs}$ & $N_{{\rm low}-z}$ & $f_{\rm contami}$ & Spec. Obs. \\
(1) & (2) & (3) & (4) & (5) \\ \hline
\multicolumn{5}{c}{ Bright (${\it NB}<24$) } \\ \hline
6.6 & 12 & 4 & 0.33 & FOCAS$^a$, LDSS3$^a$ \\ 
5.7 & 6 & 1 & 0.17 & FOCAS$^a$, LDSS3$^a$ \\ \hline
\multicolumn{5}{c}{ All } \\ \hline
6.6 & 28 & 4 & 0.14 & FOCAS$^a$, LDSS3$^{ab}$, IMACS$^c$ \\ 
5.7 & 53 & 4 & 0.08 & FOCAS$^a$, LDSS3$^a$, IMACS$^c$ \\ \hline
  \end{tabular}}\label{tab_contami}
  \begin{tabnote}
    (1) Redshift of the LAE sample. (2) Number of spectroscopically observed HSC LAEs. (3) Number of low-$z$ contaminants. (4) Contamination rates. (5) Spectroscopic follow-up observations. Only for the observations whose $N_{\rm obs}$ and $N_{{\rm low}-z}$ are found.  \\
    $^a$ This study. \\
    $^b$ Y. Harikane et al. in preparation. \\
    $^c$ R. Higuchi in preparation. \\
  \end{tabnote}
\end{table}

\subsection{Ly$\alpha$ Line Width}\label{sec_lya_width}

To quantify the Ly$\alpha$ line profiles, we measure the FWHM velocity width, $\Delta V_{\rm FWHM}$. We fit the symmetric Gaussian profile to the Ly$\alpha$ emission lines, and obtain the observed FWHM velocity width, $\Delta V_{\rm obs}$ in the same manner as that in \citet{2010ApJ...723..869O} for consistency. We correct for the instrumental broadening of line profile, and obtain $\Delta V_{\rm FWHM}$ by $\Delta V_{\rm FWHM}=\sqrt{\Delta V_{\rm obs}^2 - \Delta V_{\rm inst}^2}$, where $\Delta V_{\rm obs}$ and $\Delta V_{\rm inst}$ are FWHM velocity widths for the observed Ly$\alpha$ lines and the instrumental resolution, respectively. We use the uncertainties in the $\chi^2$ minimization fitting as the $\Delta V_{\rm FWHM}$ errors. The $\Delta V_{\rm FWHM}$ values are listed in Table \ref{tab_phys}. 

Figure \ref{fig_phys_fwhm} presents $\Delta V_{\rm FWHM}$ as a function of $L_{\rm Ly\alpha}$. We find that the bright LAEs have $\Delta V_{\rm FWHM} \simeq200-400$ km s$^{-1}$ similar to $z\simeq6$ faint LAEs with $\log L_{\rm Ly\alpha}/{\rm [erg\ s^{-1}]}\lesssim 43$ (\cite{2010ApJ...723..869O}). The narrow Ly$\alpha$ emission lines of $\Delta V\simeq200-400$ km s$^{-1}$ indicate that the bright LAEs are not broad-line AGNs. 

To quantify the relation between $\Delta V_{\rm FWHM}$ and $L_{\rm Ly\alpha}$ in Figure \ref{fig_phys_fwhm}, we carry out Spearman rank correlation tests. In this test, we find a marginal correlation at the $\simeq1.7\sigma$ significance level, possibly suggesting that $\Delta V_{\rm FWHM}$ increases with increasing $L_{\rm Ly\alpha}$.

\begin{longtable}{*{8}{c}}
\caption{Photometric Properties of Bright LAEs with Spectroscopic Redshifts}\label{tab_phot}
\hline
Object ID & $\alpha(J2000)$ & $\delta(J2000)$ & $z_{\rm Ly\alpha}$ & {\it NB} & {\it i} & {\it z} & {\it y} \\
 & & & & (mag) & (mag) & (mag) & (mag) \\
(1) & (2) & (3) & (4) & (5) & (6) & (7) & (8) \\
\hline
\endfirsthead
\endhead
\hline
\endfoot
\hline 
\multicolumn{8}{l}{(1) Object ID.} \\
\multicolumn{8}{l}{(2) Right ascension.} \\
\multicolumn{8}{l}{(3) Declination. } \\
\multicolumn{8}{l}{(4) Spectroscopic redshift of Ly$\alpha$ emission line. } \\
\multicolumn{8}{l}{(5) Total magnitudes of ${\it NB}921$ and ${\it NB}816$ for $z\simeq6.6$ and $z\simeq5.7$ LAEs, respectively. } \\
\multicolumn{8}{l}{(6)-(8) Total magnitudes of {\it i}-, {\it z}-, and {\it y}-band. } \\
\multicolumn{8}{l}{(6)-(8) $2\sigma$ limiting magnitudes for undetected bands. } \\
\multicolumn{8}{l}{$^a$ Spectroscopically confirmed with Magellan/IMACS. See Section \ref{sec_imacs}. } \\
\multicolumn{8}{l}{$^b$ COLA1 in \citet{2016ApJ...825L...7H}. } \\
\multicolumn{8}{l}{$^c$ CR7 in \citet{2015ApJ...808..139S}. } \\
\multicolumn{8}{l}{$^d$ Himiko in \citet{2009ApJ...696.1164O}. } \\
\multicolumn{8}{l}{$^e$ MASOSA in \citet{2015ApJ...808..139S}. } \\
\multicolumn{8}{l}{$^f$ Spectroscopically confirmed in \citet{2012ApJ...760..128M}. } \\
\endlastfoot
\multicolumn{8}{c}{${\it NB}921$ ($z\simeq6.6$)} \\ \hline 
HSC J162126$+$545719 & 16:21:26.51 & $+$54:57:19.14 & 6.545 & $22.33\pm0.02$ & $>25.8$ & $23.77\pm0.18$ & $22.92\pm0.16$ \\ 
HSC J233125$-$005216 & 23:31:25.36 & $-$00:52:16.36 & 6.559 & $23.17\pm0.08$ & $>26.6$ & $25.34\pm0.27$ & $24.96\pm0.37$ \\ 
HSC J160234$+$545319 & 16:02:34.77 & $+$54:53:19.95 & 6.576 & $23.24\pm0.05$ & $>26.4$ & $24.79\pm0.26$ & $>24.8$ \\ 
HSC J160940$+$541409 & 16:09:40.25 & $+$54:14:09.04 & 6.564 & $23.52\pm0.06$ & $>26.4$ & $25.45\pm0.32$ & $>24.7$ \\ 
HSC J100334$+$024546 & 10:03:34.66 & $+$02:45:46.56 & 6.575 & $23.61\pm0.05$ &$>26.7$ & $25.62\pm0.27$ & $24.97\pm0.29$ \\ 
HSC J100550$+$023401 & 10:05:50.97 & $+$02:34:01.51 & 6.573 & $23.71\pm0.10$ & $>26.4$ & $25.15\pm0.22$ & $>25.3$ \\ 
HSC J160707$+$555347 & 16:07:07.48 & $+$55:53:47.90 & 6.586 & $23.86\pm0.09$ & $>26.5$ & $25.35\pm0.32$ & $>24.8$\\ 
HSC J160107$+$550720 & 16:01:07.45 & $+$55:07:20.63 & 6.563 & $23.96\pm0.12$ & $>26.4$ & $>25.5$ & $>24.4$ \\ 
\hline 
\multicolumn{8}{c}{${\it NB}816$ ($z\simeq5.7$)} \\ \hline 
HSC J233408$+$004403 & 23:34:08.79 & $+$00:44:03.78 & 5.707 & $22.85\pm0.04$ & $25.40\pm0.20$& $>25.8$ & $>25.1$ \\ 
HSC J021835$-$042321$^{a}$ & 02:18:35.94 & $-$04:23:21.62 & 5.757 & $23.10\pm0.06$ & $25.38\pm0.22$ & $24.93\pm0.19$ & $25.23\pm0.56$ \\ 
HSC J233454$+$003603 & 23:34:54.95 & $+$00:36:03.99 & 5.732 & $23.16\pm0.05$ & $25.42\pm0.19$ & $25.60\pm0.37$ & $24.59\pm0.28$ \\ 
HSC J021752$-$053511 & 02:17:52.63 & $-$05:35:11.78 & 5.756 & $23.17\pm0.05$ & $25.24\pm0.12$ & $24.50\pm0.14$ & $24.42\pm0.20$ \\ 
HSC J021828$-$051423$^{a}$ & 02:18:28.87 & $-$05:14:23.01 & 5.737 & $23.57\pm0.04$ & $26.25\pm0.22$ & $26.27\pm0.38$ & $>25.78$ \\ 
HSC J021724$-$053309$^{a}$ & 02:17:24.02 & $-$05:33:09.61 & 5.707 & $23.64\pm0.08$ & $>25.8$ & $25.36\pm0.29$ & $>25.4$ \\ 
HSC J021859$-$052916$^{a}$ & 02:18:59.92 & $-$05:29:16.81 & 5.674 & $23.71\pm0.06$ & $25.17\pm0.14$ & $24.05\pm0.09$ & $24.00\pm0.17$ \\ 
HSC J021836$-$053528$^{a}$ & 02:18:36.37 & $-$05:35:28.07 & 5.700 & $23.75\pm0.06$ & $25.95\pm0.22$ & $25.20\pm0.21$ & $24.88\pm0.25$ \\ 
HSC J232558$+$002557 & 23:25:58.43 & $+$00:25:57.53 & 5.703 & $23.78\pm0.09$ & $25.86\pm0.22$ & $25.29\pm0.28$ & $>24.9$ \\ 
HSC J022001$-$051637 & 02:20:01.10 & $-$05:16:37.51 & 5.708 & $23.79\pm0.04$ & $26.04\pm0.19$ & $25.99\pm0.30$ & $>25.8$ \\ 
HSC J021827$-$044736$^{a}$ & 02:18:27.44 & $-$04:47:36.98 & 5.703 & $23.80\pm0.08$ & $26.93\pm0.38$ & $>26.3$ & $>25.8$ \\ 
HSC J021830$-$051457$^{a}$ & 02:18:30.53 & $-$05:14:57.81 & 5.688 & $23.83\pm0.05$ & $25.93\pm0.17$ & $26.27\pm0.38$ & $>25.8$ \\ 
HSC J021624$-$045516$^{a}$ & 02:16:24.70 & $-$04:55:16.55 & 5.706 & $23.94\pm0.06$ & $26.24\pm0.22$ & $25.67\pm0.23$ & $>25.5$ \\ 
\hline
\multicolumn{8}{c}{Previously identified bright LAEs} \\ \hline 
HSC J100235$+$021213$^{b}$ & 10:02:35.38 & $+$02:12:13.96 & 6.593 & $23.18\pm0.03$ & $>26.9$ & $24.98\pm0.12$ & $25.29\pm0.31$ \\ 
HSC J100058$+$014815$^{c}$ & 10:00:58.00 & $+$01:48:15.14 & 6.604 & $23.25\pm0.03$ & $>26.9$ & $25.12\pm0.13$ & $24.48\pm0.16$ \\ 
HSC J021757$-$050844$^{d}$ & 02:17:57.58 & $-$05:08:44.63 & 6.595 & $23.50\pm0.03$ & $>27.4$ & $25.77\pm0.20$ & $25.40\pm0.27$ \\ 
HSC J100124$+$023145$^{e}$ & 10:01:24.79 & $+$02:31:45.38 & 6.541 & $23.61\pm0.03$ & $>27.0$ & $25.25\pm0.14$ & $25.64\pm0.40$ \\ 
HSC J100109$+$021513$^{f}$ & 10:01:09.72 & $+$02:15:13.45 & 5.712 & $23.13\pm0.02$ & $25.77\pm0.13$ & $25.91\pm0.21$ & $25.97\pm0.41$  \\ 
HSC J100129$+$014929$^{f}$ & 10:01:29.07 & $+$01:49:29.81 & 5.707 & $23.47\pm0.02$ & $25.87\pm0.15$ & $25.27\pm0.13$ & $25.30\pm0.28$  \\ 
HSC J100123$+$015600$^{f}$ & 10:01:23.84 & $+$01:56:00.46 & 5.726 & $23.94\pm0.03$ & $26.43\pm0.25$ & $25.85\pm0.21$ & $>25.9$  \\ 
\hline 
\end{longtable}

\begin{longtable}{*{10}{c}}
\caption{Low-$z$ Contamination Sources}\label{tab_lowz}
\hline
Object ID & $\alpha(J2000)$ & $\delta(J2000)$ & $z_{\rm spec}$ & {\it NB} & {\it g} & {\it r} & {\it i} & {\it z} & {\it y}   \\
& & & & (mag) & (mag) & (mag) & (mag) & (mag) & (mag) \\
(1) & (2) & (3) & (4) & (5) & (6) & (7) & (8) & (9) & (10) \\
\hline
\endfirsthead
\endhead
\hline
\endfoot
\hline
\multicolumn{10}{l}{(1) Object ID.} \\
\multicolumn{10}{l}{(2) Right ascension.} \\
\multicolumn{10}{l}{(3) Declination. } \\
\multicolumn{10}{l}{(4) Spectroscopic redshift. } \\
\multicolumn{10}{l}{(5) Total magnitudes of ${\it NB}921$ and ${\it NB}816$ for $z\simeq6.6$ and $z\simeq5.7$ LAEs, respectively. } \\
\multicolumn{10}{l}{(6)-(10) Total magnitudes of {\it g}-, {\it r}-, {\it i}-, {\it z}-, and {\it y}-band. } \\
\multicolumn{10}{l}{(6)-(10) $2\sigma$ limiting magnitudes for undetected bands. } \\
\endlastfoot
\multicolumn{10}{c}{${\it NB}921$} \\ \hline 
HSC J1001$+$0229 & 10:01:44.34 & $+$02:29:09.96 & $0.840$ & 23.64 & 26.78 & $>26.7$ & $>26.5$ & 25.10 & $>25.0$  \\ 
HSC J0957$+$0306 & 09:57:16.07 & $+$03:06:30.31 & $0.841$ & 23.73 & $>27.2$ & $>26.7$ & $>26.5$ & 25.15 & $>25.0$ \\ 
HSC J1611$+$5541 & 16:11:30.34 & $+$55:41:00.39 & $0.844$  & 23.82 & $>27.2$ & $>26.7$ & 26.37 & 25.47 & $>25.0$ \\ 
HSC J1609$+$5620 & 16:09:18.03 & $+$56:20:50.89 & $0.838$ & 23.96 & $>27.2$ & $>26.7$ & $>26.5$ & 25.52 & $>25.0$ \\ 
\hline 
\multicolumn{10}{c}{${\it NB}816$} \\ \hline 
HSC J2327$+$0054 & 23:27:48.16 & $+$00:54:20.84 & $0.639$ & 23.18 & $>27.2$ & $>26.7$ & 25.31 & $>25.8$ & $>25.0$ \\ 
\hline
\end{longtable}

\begin{longtable}{*{7}{c}}
\caption{Physical Properties of Bright LAEs with Spectroscopic Redshifts}\label{tab_phys}
\hline
Object ID & $F_{\rm Ly\alpha}$ & $\log{L_{\rm Ly\alpha}}$ & $EW_{\rm 0,Ly\alpha}$ & $\Delta V_{\rm FWHM}$ & $M_{\rm UV}$ & Extended?$^{a}$ \\
& (erg s$^{-1}$ cm$^{-2}$) & (erg s$^{-1}$) & (\AA) & (km s$^{-1}$) & (mag) & \\
(1) & (2) & (3) & (4) & (5) & (6) & (7) \\
\hline
\endfirsthead
\endhead
\hline
\endfoot
\hline
\multicolumn{7}{l}{(1) Object ID. } \\
\multicolumn{7}{l}{(2) Ly$\alpha$ flux in units of $10^{-17}$ erg s$^{-1}$ cm$^{-2}$.} \\
\multicolumn{7}{l}{(3) Ly$\alpha$ luminosity. } \\
\multicolumn{7}{l}{(4) Ly$\alpha$ EW.} \\
\multicolumn{7}{l}{(5) Velocity FWHM of the Ly$\alpha$ emission line.} \\
\multicolumn{7}{l}{(6) Absolute UV magnitude.} \\
\multicolumn{7}{l}{(7) Flag of the Ly$\alpha$ spatial extent.} \\
\multicolumn{7}{l}{$^a$ If the column is {\tt Y}, the object is spatially extended in Ly$\alpha$. See \citet{2017arXiv170408140S}. } \\
\multicolumn{7}{l}{$^b$ Spectroscopically confirmed with Magellan/IMACS. See Section \ref{sec_imacs}. } \\
\multicolumn{7}{l}{$^c$ Physical quantities in the columns (2)-(5) are obtained from literature. } \\
\multicolumn{7}{l}{$^d$ COLA1 in \citet{2016ApJ...825L...7H}. } \\
\multicolumn{7}{l}{$^e$ CR7 in \citet{2015ApJ...808..139S}. } \\
\multicolumn{7}{l}{$^f$ Himiko in \citet{2009ApJ...696.1164O}. } \\
\multicolumn{7}{l}{$^g$ MASOSA in \citet{2015ApJ...808..139S}. } \\
\multicolumn{7}{l}{$^h$ Spectroscopically confirmed in \citet{2012ApJ...760..128M}. } \\
\multicolumn{7}{l}{$^i$ These values are calculated from the rest-frame UV continuum emission detected in the spectroscopic data. } \\
\endlastfoot
\multicolumn{7}{c}{${\it NB}921$ ($z\simeq6.6$)} \\ \hline 
HSC J162126$+$545719 & $16.0\pm0.12$ & $43.89\pm0.12$ & $98.6\pm32.7^i$ & $367\pm19$ & $-20.48\pm0.31^i$ & \\ 
HSC J233125$-$005216 & $8.20\pm0.05$ & $43.60\pm0.15$ & $80.8\pm33.4$ & $168\pm17$ & $-21.87\pm0.37$ & \\ 
HSC J160234$+$545319 & $6.74\pm0.03$ & $43.52\pm0.002$ & $>57.3$ & $394\pm21$ & $>-22.0$ & \\ 
HSC J160940$+$541409 & $3.98\pm0.06$ & $43.29\pm0.006$ & $>30.8$ & $302\pm29$ & $>-22.1$ & Y \\ 
HSC J100334$+$024546 & $6.14\pm0.13$ & $43.48\pm0.18$ & $61.1\pm18.9$ & $239\pm18$ & $-21.87\pm0.29$ & \\ 
HSC J100550$+$023401 & $7.94\pm0.10$ & $43.59\pm0.005$ & $>107.0$ & $312\pm34$ & $>-21.5$ & \\ 
HSC J160707$+$555347 & $6.07\pm0.05$ & $43.48\pm0.004$ & $>51.5$ & $397\pm30$ & $>-22.0$ & \\ 
HSC J160107$+$550720 & $2.45\pm0.03$ & $43.08\pm0.005$ & $>14.4$ & $393\pm33$ & $>-22.4$ & \\ 
\hline 
\multicolumn{7}{c}{${\it NB}816$ ($z\simeq5.7$)} \\ \hline 
HSC J233408$+$004403 & $13.5\pm0.03$ & $43.68\pm0.001$ & $>256.4$ & $323\pm18$ & $>-20.8$ & \\ 
HSC J021835$-$042321$^{b}$ & $12.5\pm0.07$ & $43.66\pm0.08$ & $107.4\pm21.4$ & $298\pm34$ & $-21.70\pm0.19$ & \\ 
HSC J233454$+$003603 & $13.6\pm0.10$ & $43.69\pm0.14$ & $216.6\pm88.5$ & $318\pm19$ & $-21.02\pm0.37$ & \\ 
HSC J021752$-$053511 & $12.7\pm0.09$ & $43.66\pm0.06$ & $73.5\pm10.7$ & $157\pm8$ & $-22.13\pm0.14$ & \\ 
HSC J021828$-$051423$^{b}$ & $7.04\pm0.06$ & $43.40\pm0.15$ & $207.3\pm87.2$ & $<410$ & $-20.35\pm0.38$ & \\ 
HSC J021724$-$053309$^{b}$  & $5.48\pm0.02$ & $43.29\pm0.12$ & $69.5\pm21.9$ & $<410$ & $-21.25\pm0.29$ & \\ 
HSC J021859$-$052916$^{b}$ & $4.55\pm0.05$ & $43.20\pm0.04$ & $17.2\pm1.8$ & $311\pm44$ & $-22.55\pm0.09$ & \\ 
HSC J021836$-$053528$^{b}$ & $4.90\pm0.03$ & $43.24\pm0.09$ & $53.6\pm11.8$ & $<410$ & $-21.41\pm0.21$ & \\ 
HSC J232558$+$002557 & $3.59\pm0.02$ & $43.10\pm0.12$ & $42.7\pm13.1$ & $373\pm31$ & $-21.32\pm0.28$ & \\ 
HSC J022001$-$051637 & $5.10\pm0.03$ & $43.26\pm0.12$ & $115.6\pm37.1$ & $271\pm30$ & $-20.62\pm0.30$ & \\ 
HSC J021827$-$044736$^{b}$ & $5.34\pm0.05$ & $43.28\pm0.03$ & $>160.8$ & $<410$ & $>-20.3$ & \\ 
HSC J021830$-$051457$^{b}$ & $7.19\pm0.13$ & $43.40\pm0.15$ & $210.3\pm88.8$ & $<410$ & $-20.34\pm0.38$ & \\ 
HSC J021624$-$045516$^{b}$ & $4.17\pm0.03$ & $43.17\pm0.09$ & $70.5\pm17.1$ & $<410$ & $-20.94\pm0.23$ & \\ 
\hline
\multicolumn{7}{c}{Previously identified bright LAEs$^{c}$} \\ \hline 
HSC J100235$+$021213$^{d}$ & $16.0$ & 43.9 & 53 & 194 & $-21.55\pm0.31$ & \\ 
HSC J100058$+$014815$^{e}$ & $12.7\pm0.08$ & 43.8 & 211 & 266 & $-22.37\pm0.16$ & Y \\ 
HSC J021757$-$050844$^{f}$ & $5.06\pm0.32$ & 43.4 & 78 & 251 & $-21.44\pm0.27$ & Y \\ 
HSC J100124$+$023145$^{g}$ & $5.16$ & 43.4 & $>206$ & 386 & $-21.19\pm0.40$ & \\ 
HSC J100109$+$021513$^{h}$ & $7.32\pm0.85$ & $43.4$ & $19.7^{+9.00}_{-7.93}$ & $265\pm71$ & $-20.64\pm0.41$ & Y \\ 
HSC J100129$+$014929$^{h}$ & $5.77\pm0.61$ & $43.3$ & $60.9^{+5.89}_{-41.32}$ & $422\pm120$ & $-21.31\pm0.28$ & Y \\ 
HSC J100123$+$015600$^{h}$ & $3.79\pm0.66$ & $43.1$ & $11.4^{+8.01}_{-7.32}$ & $237\pm58$ & $>-20.72$ & \\ 
\hline
\end{longtable}

\begin{longtable}{*{9}{c}}
\caption{UV Nebular Emission Lines of Bright LAEs}\label{tab_uvline}
\hline
Object ID & \multicolumn{4}{c}{Flux ($EW_0$) ($2\sigma$ upper limits)} & \multicolumn{4}{c}{Line flux ratio relative to Ly$\alpha$}   \\
(R.A.) & N {\sc v} & C {\sc iv} & He {\sc ii} & O {\sc iii}] & N {\sc v} & C {\sc iv} & He {\sc ii} & O {\sc iii}]  \\
& \multicolumn{4}{c}{} & /Ly$\alpha$ & /Ly$\alpha$ & /Ly$\alpha$ & /Ly$\alpha$ \\
& \multicolumn{4}{c}{($10^{-17}$ erg s$^{-1}$ cm$^{-2}$) (\AA)} & & & & \\
(1) & (2) & (3) & (4) & (5) & (6) & (7) & (8) & (9) \\
\hline
\endfirsthead
\endhead
\hline
\endfoot
\hline
\multicolumn{9}{l}{(1) Object ID.} \\ 
\multicolumn{9}{l}{(2)-(5) Flux and $2\sigma$ flux upper limits of the C {\sc iv}, He {\sc ii}, and O {\sc iii}] emission lines. } \\
\multicolumn{9}{l}{The number in the parentheses is the EW and $2\sigma$ limits of the C {\sc iv}, He {\sc ii}, and O {\sc iii}] emission lines. } \\
\multicolumn{9}{l}{(6)-(9) Line flux ratios of the UV-nebular emission lines relative to Ly$\alpha$. } \\
\endlastfoot
J162126 & $<0.81 (<7.2)$ & $<0.13 (<1.8)$ & $<0.35 (<5.4)$ & $<0.22 (<3.5)$ & $<0.05$ & $<0.01$ & $<0.02$ & $<0.01$ \\ 
J233125 & $<0.53$ & ... & ... & ... & $<0.06$ & ... & ... & ... \\ 
J160234 & $<0.71$ & $<0.81$ & $<1.06$ & $<1.55$ & $<0.11$ & $<0.12$ & $<0.16$ & $<0.23$ \\ 
J160940 & $<0.56$ & $<0.77$ & $<1.20$ & $<1.95$ & $<0.14$ & $<0.19$ & $<0.30$ & $<0.49$ \\ 
J100334 & $<0.75$ & ... & ... & ... & $<0.12$ & ... & ... & ... \\ 
J100550 & $<0.64 (<6.6)$ & $<0.07 (<1.1)$ & $<0.05 (0.91)$ & $<0.16 (<3.0)$ & $<0.08$ & $<0.01$ & $<0.01$ & $<0.03$ \\ 
J160707 & $<0.55$ & ... & ... & ... & $<0.09$ & ... & ... & ... \\ 
J160107 & $<0.66$ & ... & ... & ... & $<0.27$ & ... & ... & ... \\ 
\hline 
J233408 & $<0.67$ & $1.15 (>42)$ & $<0.16$ & $<0.09$ & $<0.05$ & $0.08\pm0.008$ & $<0.01$ & $<0.01$ \\ 
J021835 & $<0.92 (<8.2)$ & $<0.30 (<4.2)$ & $<0.39 (<6.1)$ & $<0.06 (<1.0)$ & $<0.07$ & $<0.02$ & $<0.03$ & $<0.01$ \\ 
J233454 & $<0.64 (<11)$ & $<0.08 (<2.1)$ & $<0.12 (<3.5)$ & $<0.13 (<3.9)$ & $<0.05$ & $<0.01$ & $<0.01$ & $<0.01$ \\ 
J201752 & $<0.70$ & ... & ... & ... & $<0.06$ & ... & ... & ... \\ 
J021828 & $<0.62$ & ... & ... & ... & $<0.09$ & ... & ... & ... \\ 
J021724 & $<0.15$ & ... & ... & ... & $<0.03$ & ... & ... & ... \\ 
J021859 & $<0.59$ & ... & ... & ... & $<0.13$ & ... & ... & ... \\ 
J021836 & $<0.26$ & ... & ... & ... & $<0.05$ & ... & ... & ... \\ 
J232558 & $<0.45$ & ... & ... & ... & $<0.13$ & ... & ... & ... \\ 
J022001 & $<0.72$ & ... & ... & ... & $<0.14$ & ... & ... & ... \\ 
J021827 & $<1.05$ & ... & ... & ... & $<0.20$ & ... & ... & ... \\ 
J021830 & $<1.24$ & ... & ... & ... & $<0.17$ & ... & ... & ... \\ 
J021624 & $<0.43$ & ... & ... & ... & $<0.10$ & ... & ... & ... \\ 
\hline
\end{longtable}

\begin{figure*}[t!]
 \begin{center}
  \includegraphics[width=165mm]{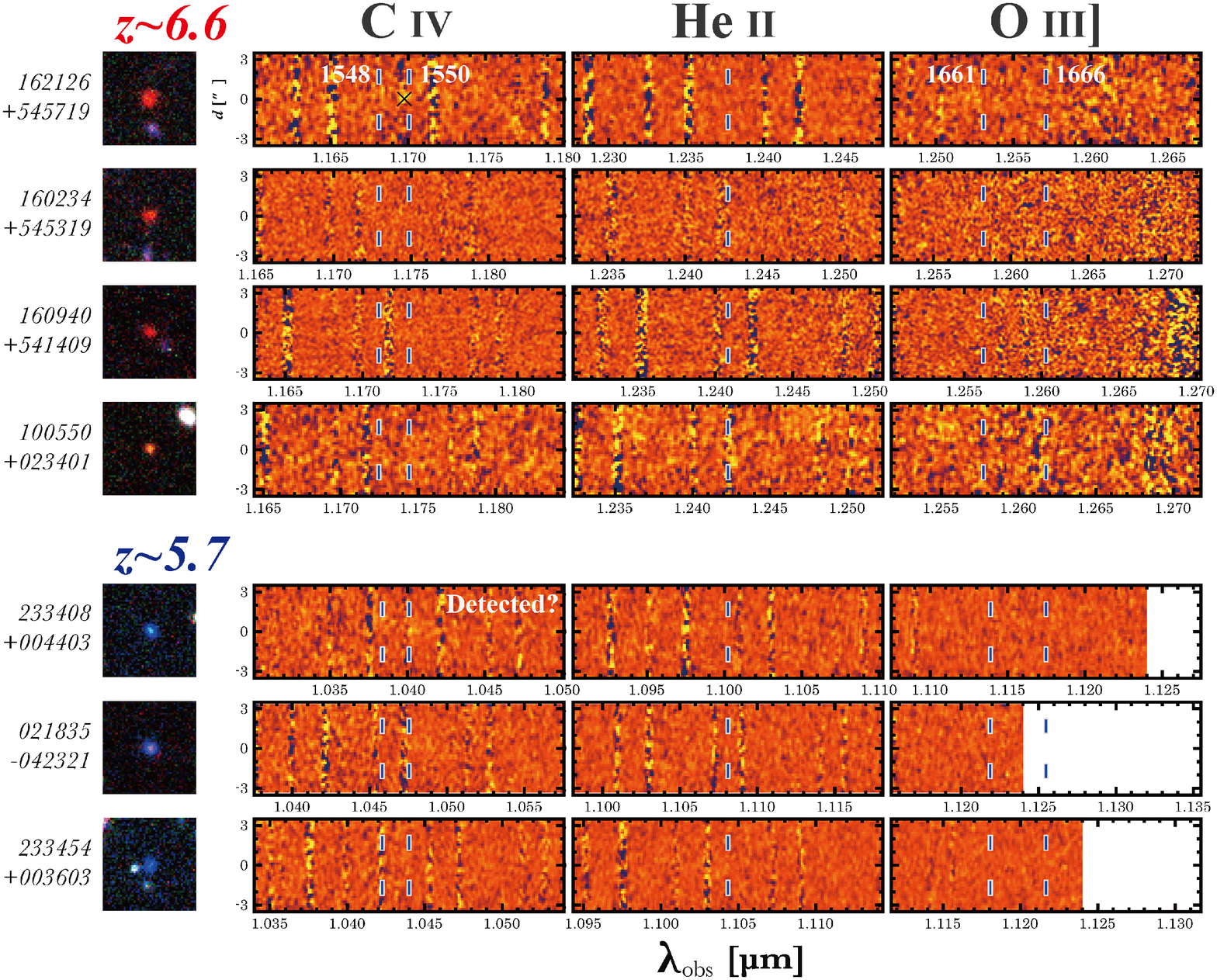}
 \end{center}
 \caption{NIR spectra for the bright LAEs at $z\simeq6.6$ (the upper four spectra) and $\simeq5.7$ (the lower three spectra). The left figures indicate the three-color composite images of the bright LAEs. The blue ticks denote the C {\sc iv} (left), He {\sc ii} (center), and O {\sc iii}] (right) wavelengths which are expected from the redshift of Ly$\alpha$ emission line. For HSC J162126$+$545719, the emission feature near the expected C {\sc iv}$\lambda1550$ wavelength is likely to be a residual of the sky subtraction, which is marked by a black cross. A C {\sc iv}$\lambda1550$ emission line is tentatively detected in the spectrum of HSC J233408$+$004403 (see Figure \ref{fig_spec_deep1_civ}), which is discussed in Section \ref{sec_civ}. }\label{fig_hsc_spec_all_nir}
\end{figure*}

\subsection{X-ray, Mid-IR, and Radio Detectability}\label{sec_x_radio}

We check X-ray, mid-IR (MIR), and radio data to investigate whether the bright LAEs have a signature of AGN activities. Such X-ray, MIR, and radio data are available in the UD fields, UD-COSMOS and UD-SXDS. In UD-COSMOS, an object (i.e. HSC J100334$+$024546) is covered by MIR and radio data. In UD-SXDS, all the 10 objects are observed in X-ray, MIR, and radio. For the X-ray data, we use the {\it XMM-Newton} source catalog whose sensitivity limit is $f_{\rm 0.5-2 keV}=6\times 10^{-16}$ erg cm$^{-2}$ s$^{-1}$ (\cite{2008ApJS..179..124U}). For the MIR data, we use {\it Spitzer}/MIPS $24\mu m$ source catalogs for UD-COSMOS (\cite{2007ApJS..172...86S}) and UD-SXDS (the SpUDS survey, PI: J. Dunlop). These {\it Spitzer}/MIPS $24\mu m$ images reach $5\sigma$ sensitivity limits of 21.2 mag and 18.0 mag in UD-COSMOS and UD-SXDS, respectively. For the radio data, we check the Very Large Array (VLA) 1.4 GHz source catalogs of \citet{2007ApJS..172...46S} for UD-COSMOS, and \citet{2006MNRAS.372..741S} for UD-SXDS. The typical r.m.s noise level of the VLA data is $f_{\rm 1.4GHz}\simeq10\mu$Jy beam$^{-1}$. 

We find that there are no counterparts in the X-ray, MIR, and radio data. The no X-ray, MIR, and radio counterparts indicate that there is no clear signature of AGN activities based on the multi-wavelength data. By considering the typical SED of AGNs (e.g., \cite{1994ApJS...95....1E, 2002ApJ...565..773T, 2003AJ....126.1131R}), the rest-frame UV luminosity of LAEs, and the sensitivity limits of these multi-wavelength data, we rule out, at least, the possibility that the LAEs have radio-loud AGNs.

\begin{figure}[t!]
 \begin{center}
  \includegraphics[width=85mm]{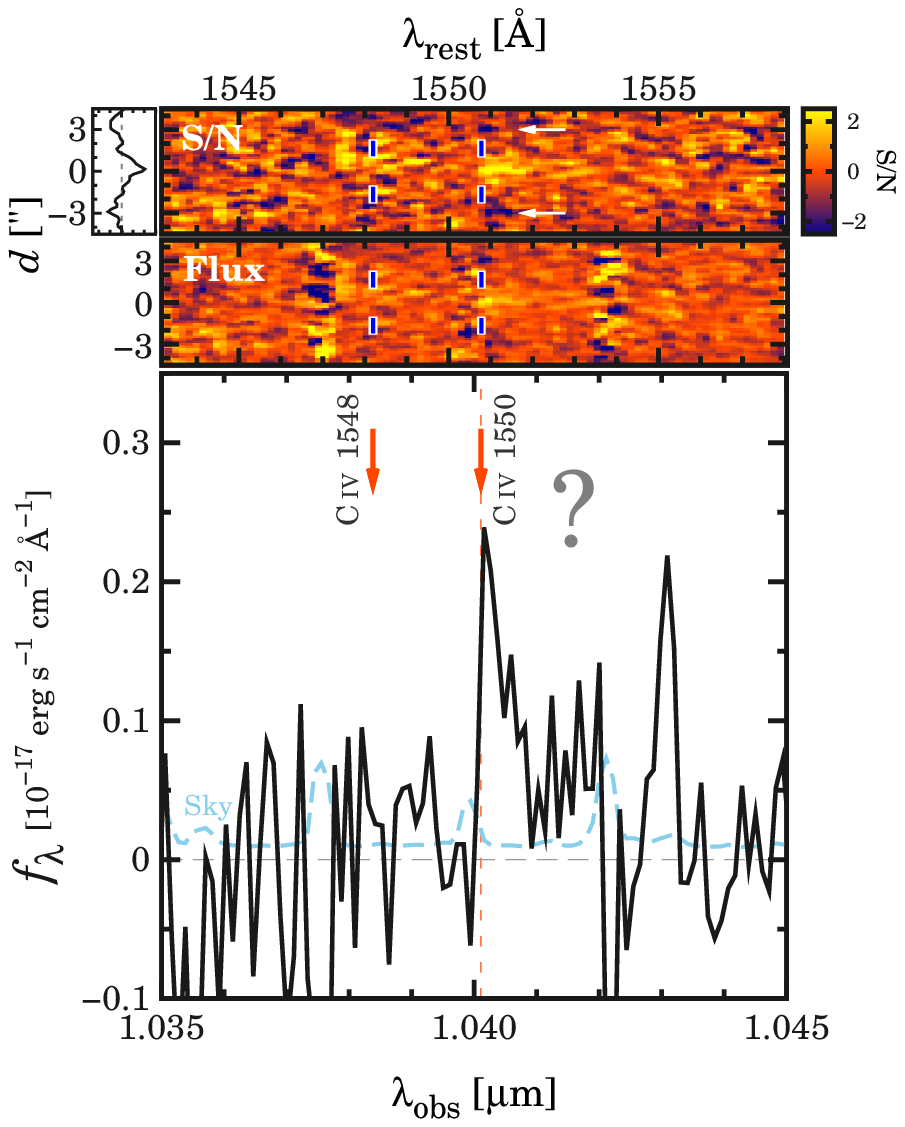}
 \end{center}
 \caption{Tentative detection of a C {\sc iv}$\lambda1550$ emission line for HSC J233408$+$004403. The 2D spectra in the top and middle panels present the $S/N$ and flux maps, respectively. The bottom panel shows the 1D spectrum of the flux map. The white arrows indicate the expected positions of the negative C {\sc iv}$\lambda1550$ emission lines which are produced in the $\pm3^{\prime\prime}$ slit dithering processes. The top-left panel depicts the 1D $S/N$ spectrum along the spatial direction at the tentative C {\sc iv}$\lambda1550$ emission line. The blue ticks and the red arrows indicate the C {\sc iv}$\lambda1548$ and C {\sc iv}$\lambda1550$ wavelengths expected from the Ly$\alpha$ emission line (i.e. $z_{\rm Ly\alpha}=5.707$). The cyan dashed line present the OH sky emission. The C {\sc iv}$\lambda1550$ emission line is tentatively detected at a significance level of $\simeq4-9$. }\label{fig_spec_deep1_civ}
\end{figure}

\begin{figure*}[t!]
 \begin{center}
  \includegraphics[width=140mm]{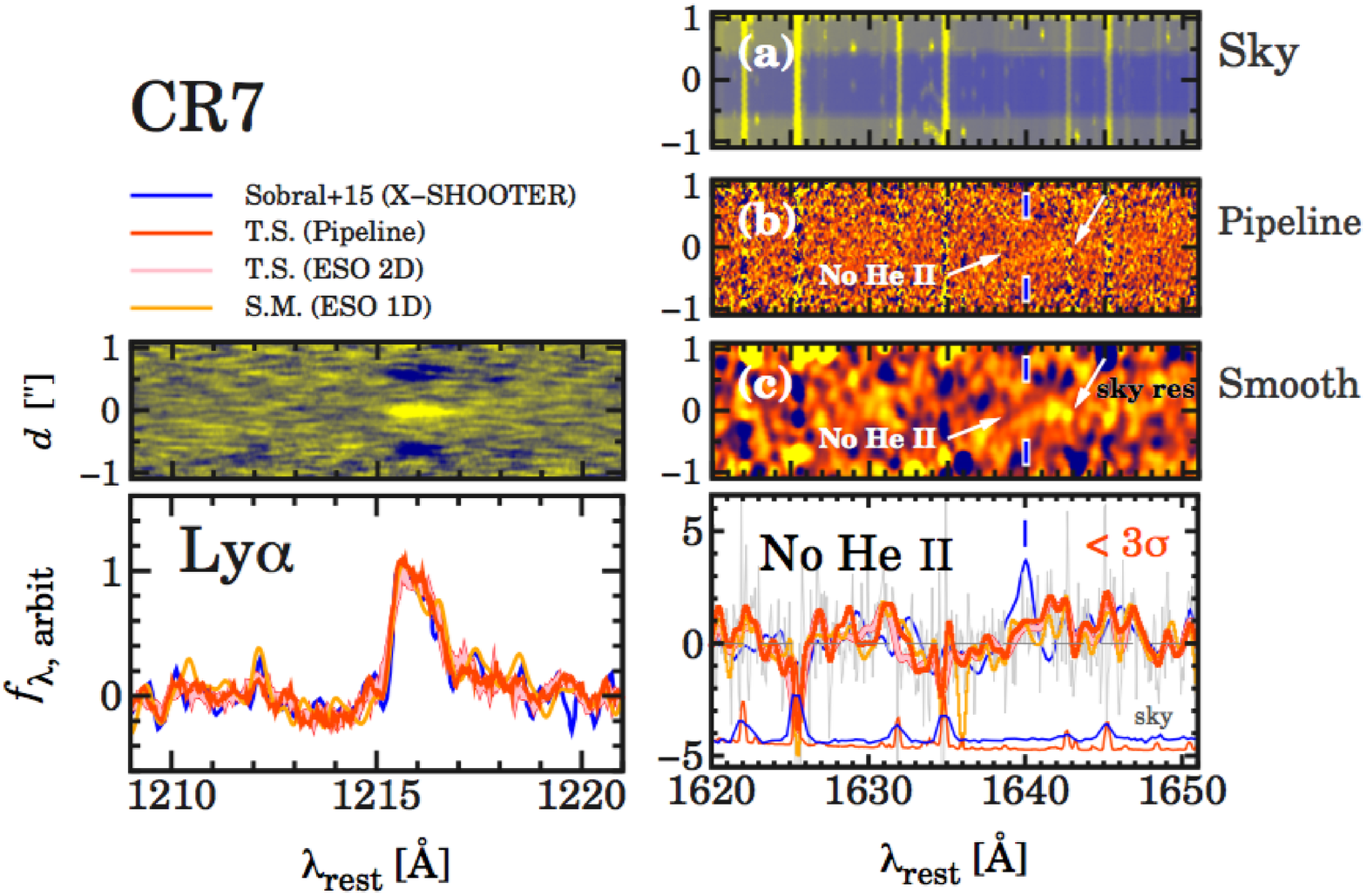}
 \end{center}
 \caption{Re-analyzed VLT/X-SHOOTER spectra of CR7. The left and right panels denote the VIS and NIR arms of the X-SHOOTER spectra. The blue lines indicate the X-SHOOTER spectra in \citet{2015ApJ...808..139S}. The red, magenta, and orange lines depict spectra obtained from (1) reducing the raw data with the X-SHOOTER reduction pipeline {\tt ESO REFLEX} ({\it Pipeline}), (2) stacking of each 2D 1-exposure spectrum reduced by ESO ({\it ESO 2D}), and (3) stacking of each 1D 1-exposure spectrum reduced by ESO ({\it ESO 1D}), respectively. These lines have been smoothed with a kernel of $\simeq0.4$\,\AA\ width which is similar to that of \citet{2015ApJ...808..139S}. The gray lines present the un-smoothed spectrum obtained from our data reduction with {\tt ESO REFLEX}. The thin blue and red lines indicate sky OH lines in \citet{2015ApJ...808..139S} and our re-analyzed data, respectively. The top-left panels show the 2D spectrum of the X-SHOOTER VIS arm. The top-right panels indicate (a) sky OH line, (b) un-smoothed and (c) smoothed 2D spectra, all of which are obtained from our data reduction with {\tt ESO REFLEX}. The feature at $\lambda_{\rm rest}=1643$\,\AA\, appears to be made by sky subtraction residuals. The blue ticks indicate the position of He {\sc ii} whose detection is claimed by \citet{2015ApJ...808..139S}. See Section \ref{sec_cr7} for more details. }\label{fig_cr7}
\end{figure*}

\subsection{UV Nebular Line Flux}\label{sec_fratio}

Here we investigate whether the rest-frame UV-nebular lines of N {\sc v}$\lambda\lambda 1238, 1240$, C {\sc iv}$\lambda\lambda1548, 1550$, He {\sc ii}$\lambda1640$, and O {\sc iii}]$\lambda\lambda1661,1666$ are detected from the bright LAEs. First, we check the detectability of N {\sc v} emission line which is a coarse indicator of AGN presence. The wavelengths of N {\sc v} are covered by the FOCAS, LDSS3, and IMACS optical spectra for both of the $z\simeq6.6$ and $z\simeq5.7$ LAE samples. In order to estimate the flux limits, we sample the 1D spectra in $\simeq10$\,\AA\ bins (comparable to the Ly$\alpha$ line FWHM) around the expected wavelengths of N {\sc v}. We then obtain the flux limit by using the flux distribution over a $\pm50$\,\AA\ range of the expected wavelengths of N {\sc v}. We find that there are no N {\sc v} emission lines for all the 21 bright LAEs. The $2\sigma$ flux limits for the N {\sc v} emission lines are listed in Table \ref{tab_uvline}. The line flux ratio of N {\sc v} to Ly$\alpha$ is typically $f_{\rm NV}/f_{\rm Ly\alpha}\lesssim10$\%. 

Next, we search for the UV-nebular emission lines of C {\sc iv}, He {\sc ii}, and O {\sc iii}] for the seven bright LAEs whose NIR spectra are obtained (Section \ref{sec_obs_nir}). Figure \ref{fig_hsc_spec_all_nir} presents the NIR spectra for the seven LAEs. Even in the deep NIR spectra with a $3\sigma$ line flux sensitivity limit of $\simeq2\times10^{-18}$ erg s$^{-1}$ cm$^{-2}$, we find no significant emission features at the expected wavelengths of redshifted He {\sc ii}, C {\sc iv}, and O {\sc iii}] lines, except for a tentative C {\sc iv} detection from a $z\simeq5.7$ LAE, HSC J233408$+$004403 (see below in this section). The flux limits for the C {\sc iv}, He {\sc ii}, and O {\sc iii}] emission lines are estimated in the same manner as that for N {\sc v}. To estimate the detection limits, we assume a single emission line even for the C {\sc iv} and O {\sc iii}] doublets which are resolved in the spectral resolution of MOSFIRE and nuMOIRCS. The $2\sigma$ flux limits for individual UV-nebular emission lines are listed in Table \ref{tab_uvline}. 

Our deep NIR spectroscopy indicates that there are no significant detections of the UV-nebular emission lines for bright LAEs. By our visual inspections for the NIR spectra, we find a tentative detection of C {\sc iv}$\lambda1550$ emission line from the brightest LAE in the $z\simeq5.7$ sample, HSC J233408$+$004403. Figure \ref{fig_spec_deep1_civ} shows the NIR spectra around the wavelengths of the C {\sc iv} emission line doublet for HSC J233408$+$004403. The C {\sc iv}$\lambda1550$ emission line is tentatively detected at the $\simeq4-9\sigma$ significance level. The significance of the line detection depends on the wavelength range of flux integration. We also identify two negative C {\sc iv}$\lambda1550$ emission lines which could be originated from the $\pm3^{\prime\prime}$ slit dithering processes in the MOSFIRE observation. Moreover, the tentative C {\sc iv}$\lambda1550$ detection might explain a possible magnitude excess in the $y$-band covering the C {\sc iv} wavelength (see Figure \ref{fig_postage_all}). The line flux is $\simeq1.2\times10^{-17}$ erg cm$^{-2}$ s$^{-1}$. The emission line has a velocity width of $\Delta V_{\rm FWHM} \simeq 50$ km s$^{-1}$ which is marginally resolved in the MOSFIRE spectral resolution. We do not detect the C {\sc iv}$\lambda$1548 component of the C {\sc iv} doublet from HSC J233408$+$004403. The single C {\sc iv} emission line at $\lambda_{\rm rest}\simeq1550$\,\AA\ may be formed by a combination of absorption and emission lines that could be originated from stellar winds and ISM. Such a C {\sc iv} line profile has been found for $z\simeq1-3$ galaxies (e.g., \cite{2003ApJ...588...65S, 2010ApJ...719.1168E, 2016ApJ...829...64D}). We discuss the emission line properties of the C {\sc iv} emitter in Section \ref{sec_civ}.

\begin{figure}[t!]
 \begin{center}
  \includegraphics[width=80mm]{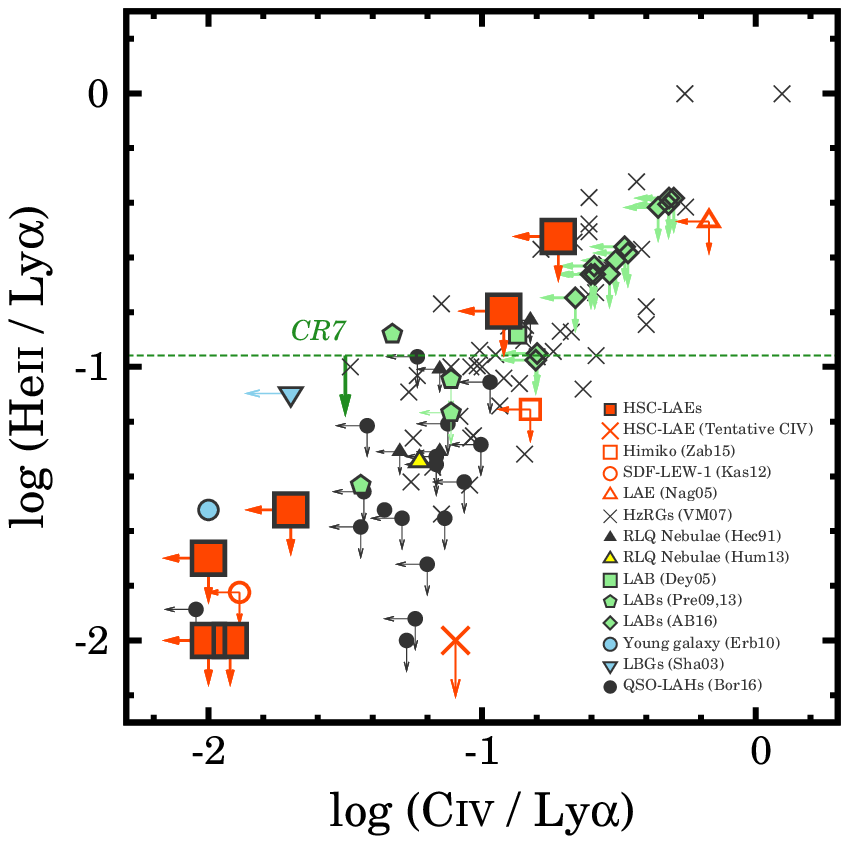} \mbox{}
 \end{center}
 \caption{Flux ratios of UV-nebular emission lines, He {\sc ii}/Ly$\alpha$ vs C {\sc iv}/Ly$\alpha$. The red filled squares indicate our bright LAEs. The red cross denotes our bright LAE whose C {\sc iv} emission line is tentatively detected (see Section \ref{sec_civ}). The red open symbols indicate $z\gtrsim6$ LAEs (red open square: Himiko in \cite{2015MNRAS.451.2050Z}; red open circle: SDF-LEW-1 in \cite{2012ApJ...761...85K}; SDF J132440.6$+$273607: \cite{2005ApJ...631L...5N}). The green arrow and dashed line are our He {\sc ii}/Ly$\alpha$ constraint for CR7 (see Section \ref{sec_cr7}). The green filled symbols represent LABs (green filled square: \cite{2005ApJ...629..654D}; green filled pentagons: \cite{2009ApJ...702..554P, 2013ApJ...762...38P}; green filled diamonds: \cite{2015ApJ...809..163A}). The cyan filled symbols present $z\simeq2-3$ star-forming galaxies (cyan filled inverse-triangle: \cite{2003ApJ...588...65S}; cyan filled circle \cite{2010ApJ...719.1168E}). The black and yellow symbols indicate AGNs, QSOs, and radio galaxies (black crosses: \cite{2007MNRAS.375.1299V}; black filled triangles: \cite{1991ApJ...381..373H}; yellow filled triangles: \cite{2013MNRAS.428..563H}; black filled circles: \cite{2016ApJ...831...39B}). The data points without a UV-nebular line detection indicate $2\sigma$ upper limits.}\label{fig_line_ratio_uv_lya}
\end{figure}

\subsection{Re-analysis of CR7 Spectra}\label{sec_cr7}

We investigate the VLT/X-SHOOTER spectrum of CR7 whose $6 \sigma$ detection of He {\sc ii} is claimed by \citet{2015ApJ...808..139S}. Two individuals of the authors in this paper and the ESO-archive service re-analyze the VLT/X-SHOOTER data that are used in the study of \citet{2015ApJ...808..139S}. We apply three methods to our re-analysis: (1) reducing the raw data with the X-SHOOTER reduction pipeline {\tt ESO REFLEX} ({\it Pipeline}), (2) stacking of each 2D 1-exposure spectrum reduced by ESO ({\it ESO 2D}), and (3) stacking of each 1D 1-exposure spectrum reduced by ESO ({\it ESO 1D}). We smooth our reduced X-SHOOTER spectra with a kernel of $\simeq0.4$\,\AA\, width which corresponds to that of \citet{2015ApJ...808..139S}. 

Figure \ref{fig_cr7} presents our reduced X-SHOOTER data of the optical (the left panel) and NIR (the right panel) arms for CR7 with the 1D spectrum obtained by \citet{2015ApJ...808..139S}. As shown in the left panel of Figure \ref{fig_cr7}, we clearly identify a Ly$\alpha$ emission line at $\lambda_{\rm rest}=1216$\,\AA. The Ly$\alpha$ line profiles of our data are in good agreement with that of the \authorcite{2015ApJ...808..139S}'s optical spectrum. However, we find no signal at $\lambda_{\rm rest}=1640$\,\AA\, where \citet{2015ApJ...808..139S} find the emission line feature (the right panel of Figure \ref{fig_cr7}). The detection significance is $<1\sigma$ at $\lambda_{\rm rest}=1640$\,\AA\, in our NIR spectra. Instead, our NIR spectra show a feature of two possible peaks at $\lambda_{\rm rest}=1643$\,\AA\, which is redder than the He {\sc ii} wavelength of \citet{2015ApJ...808..139S} by $\Delta \lambda_{\rm rest}\simeq3$\,\AA\, corresponding to the redshift difference of $\Delta z=0.01$. If we regard the two possible peaks as He {\sc ii}, we obtain a detection significance of $\simeq1.8\sigma$. This significance value is inconsistent with the $6 \sigma$ detection of \citet{2015ApJ...808..139S}. Moreover, the red component of the two possible peaks appears to be made by sky subtraction residuals, as shown in the panel (a) of Figure \ref{fig_cr7}. In the case that this red component is masked for the line flux calculation, the detection significance decreases to $\simeq1.1\sigma$. To obtain all the values of detection significance and noise levels, we use OH sky line-free regions. 

In our careful re-analysis for the X-SHOOTER data, we find that no He {\sc ii} signal of CR7 is detected. The no He {\sc ii} detection supports weak UV-nebular lines of the bright LAEs even for CR7. Based on our $S/N$-based re-analysis and the flux error from \citet{2015ApJ...808..139S}, we obtain the $3\sigma$ upper limits of He {\sc ii} flux and EW for CR7, $f_{\rm HeII}<2.1\times10^{-17}$ erg s$^{-1}$ cm$^{-2}$ and $EW_{\rm HeII}< 60$\,\AA, respectively. \footnote{Recently, the He {\sc ii}/Ly$\alpha$ line flux ratio for CR7 has been updated based on the flux recalibration of the X-SHOOTER spectrum in \citet{2017arXiv170606591M} and \citet{sobral2017}.}

\begin{figure}[h!]
 \begin{center}
  \includegraphics[width=62mm]{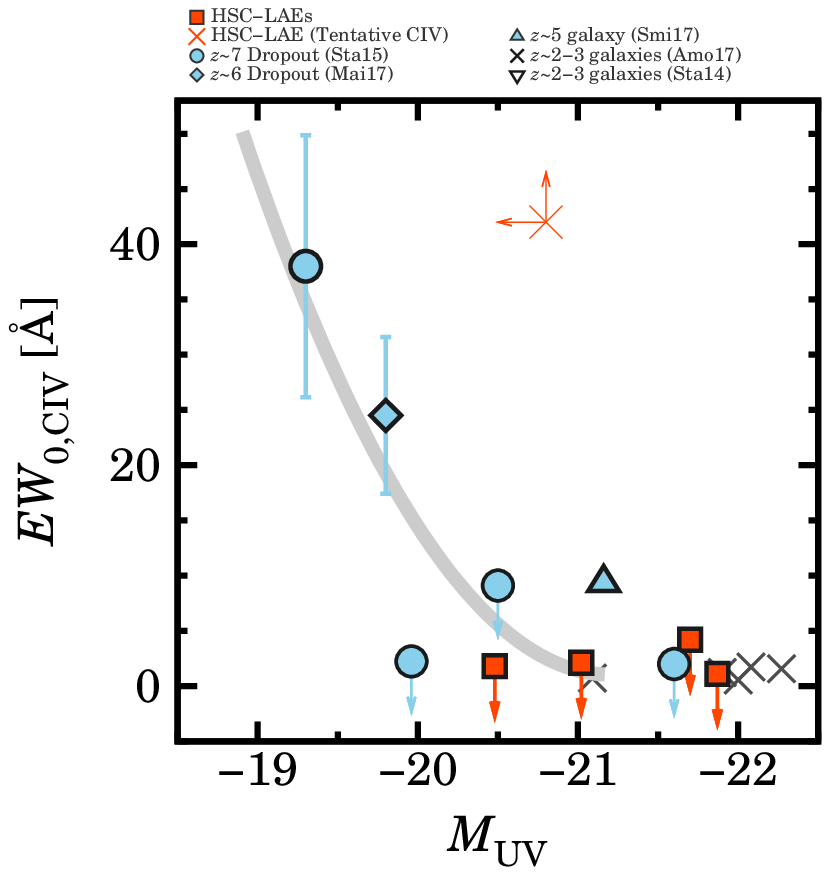}
  \includegraphics[width=62mm]{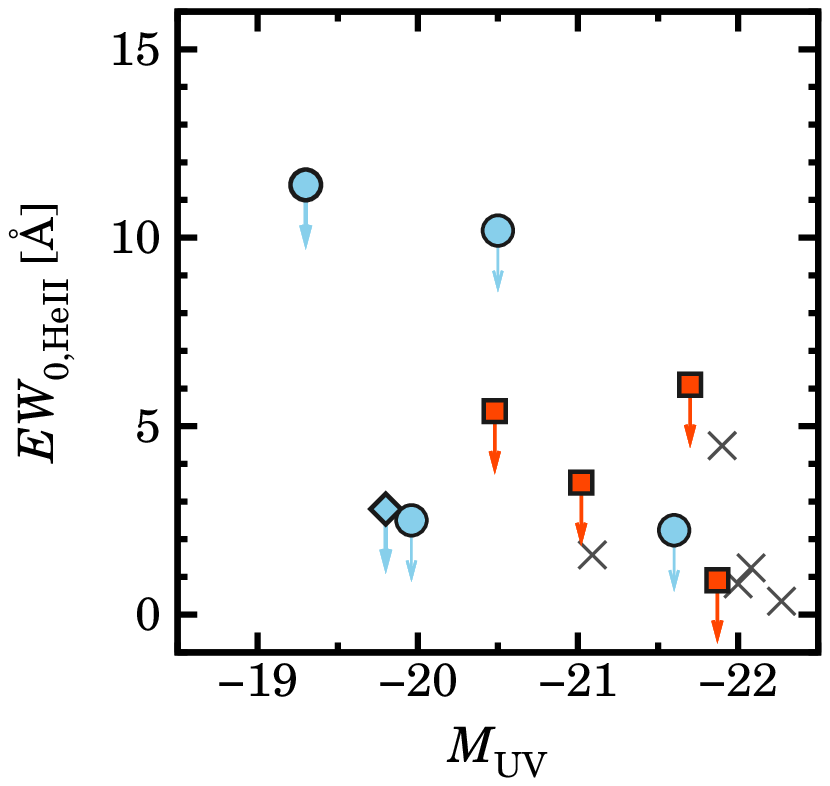}
  \includegraphics[width=62mm]{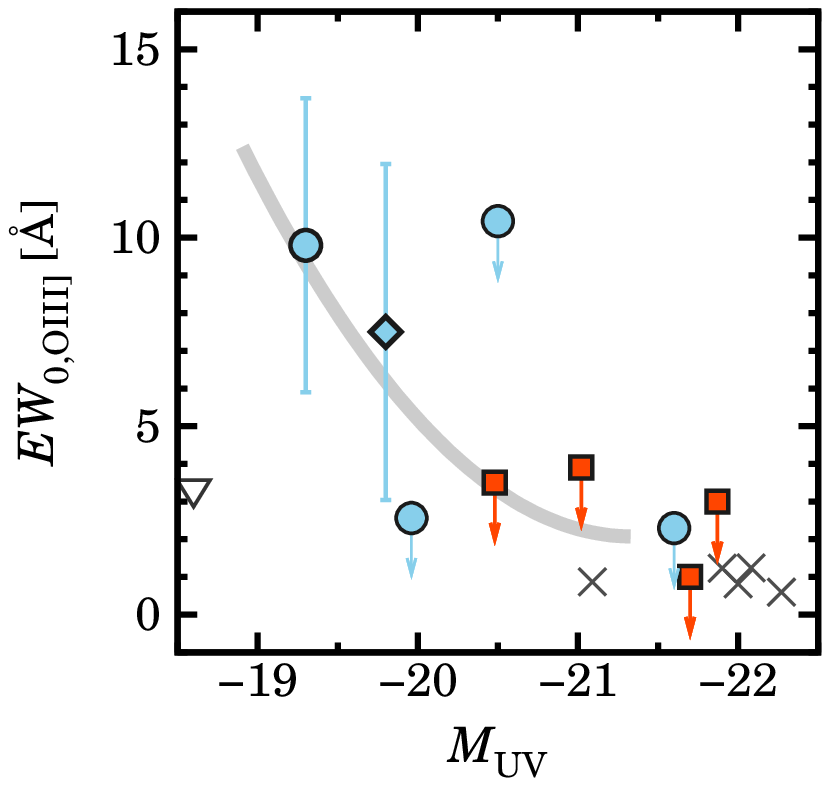}
 \end{center}
 \caption{Line equivalent widths of C {\sc iv} (top), He {\sc ii} (middle), and O {\sc iii}] (bottom) as a function of UV magnitude, $M_{\rm UV}$. The red filled squares denote our four bright LAEs with an upper limit of UV-nebular line EW. The red cross represents the LAEs whose C {\sc iv} emission is tentatively detected. The cyan filled symbols denote high-$z$ dropout galaxies (cyan filled circles: $z\simeq7$ dropouts in \citet{2015MNRAS.454.1393S}}; cyan filled diamond: $z\simeq6$ dropout in \citet{2017ApJ...836L..14M}; cyan filled triangle: \cite{2017MNRAS.467.3306S}). The gray symbols indicate $z\simeq2-3$ galaxies (gray crosses: \cite{2017NatAs...1E..52A}; gray open inverse-triangle: \cite{2014MNRAS.445.3200S}). The gray curves represent the best-fit quadratic functions to the data points of $z\simeq6-7$ dropouts in \cite{2014MNRAS.445.3200S} and \citet{2017ApJ...836L..14M} and our LAEs. The data points without a UV-nebular line detection indicate $2\sigma$ upper limits.\label{fig_muv_ewciv}
\end{figure}

\subsection{Line Flux Ratios}\label{sec_uvlya}

Figure \ref{fig_line_ratio_uv_lya} represents the line flux ratios of He {\sc ii}/Ly$\alpha$ and C {\sc iv}/Ly$\alpha$ for our bright LAEs and several Ly$\alpha$-emitting populations such as $z\simeq6-7$ LAEs (\cite{2015MNRAS.451.2050Z, 2012ApJ...761...85K, 2005ApJ...631L...5N}), spatially extended Ly$\alpha$ blobs (LABs; \cite{2005ApJ...629..654D, 2009ApJ...702..554P, 2013ApJ...762...38P, 2015ApJ...809..163A}), $z\simeq2-3$ metal-poor and star-forming galaxies (\cite{2003ApJ...588...65S, 2010ApJ...719.1168E}), AGNs, QSOs, and radio galaxies (\cite{2007MNRAS.375.1299V, 1991ApJ...381..373H, 2013MNRAS.428..563H, 2016ApJ...831...39B}).  \footnote{Note that the C {\sc iv} doublet is not spectroscopically resolved for some of the previous studies. The flux upper limit for such an unresolved C {\sc iv} doublet would be higher than that of resolved C {\sc iv} lines. But, the systematics of the C {\sc iv} flux upper limits are as small as $\simeq0.15$ dex in a flux ratio of $\log{({\rm Ly\alpha/C IV})}$, which could not affect the main conclusion. } We add CR7 with our updated He {\sc ii}/Ly$\alpha$ constraint in Section \ref{sec_cr7}. The UV-nebular lines of C {\sc iv}, He {\sc ii}, and O {\sc iii}] are not detected from all of our 7 bright LAEs even for CR7 except for a tentative C {\sc iv} detection (Section \ref{sec_civ}). Albeit with only upper limits on the line flux ratios, we find that our bright LAEs typically have flux ratios of He {\sc ii}/Ly$\alpha$ and C {\sc iv}/Ly$\alpha$ lower than those of AGNs, QSOs, radio galaxies, and LABs, but similar to those of star forming galaxies in \citet{2003ApJ...588...65S} and \citet{2010ApJ...719.1168E}. Interestingly, the UV-nebular lines are extremely faint for several of our bright LAEs. For such objects, the flux ratio of the UV-nebular lines relative to Ly$\alpha$, i.e. $f_{\rm UV\,line}/f_{\rm Ly\alpha}$, is below the order of 1 \%.

\section{Discussion}\label{sec_discuss}

\subsection{Properties of Bright LAEs at $z\simeq6$}\label{sec_nature}

We summarize the properties of the bright $z\simeq6-7$ LAEs which have been revealed in our statistical study (Section \ref{sec_results}). 
\mbox{}\\

\begin{itemize} 
\item The Ly$\alpha$ equivalent widths, $EW_{\rm 0,\, Ly\alpha}$, range from $\simeq10$\,\AA\, to $\simeq300$\,\AA. 

\item The Ly$\alpha$ line widths are $\simeq200-400$ km s$^{-1}$. 

\item There are no detections of X-ray, MIR, and radio emission.

\item The N {\sc v} emission line is not detected down to a N {\sc v}/Ly$\alpha$ flux ratio of $\simeq10$\%. 

\item Most of the bright LAEs have the compact Ly$\alpha$ emission. Only 5 objects out of the 28 bright LAEs show Ly$\alpha$ emission which are significantly extended compared to the PSF FWHM size of $\simeq0.\!\!^{\prime\prime}7$ in the ground-based HSC NB images. 

\item The UV-nebular lines of C {\sc iv}, He {\sc ii}, and O {\sc iii}] are not detected from all of our 7 bright LAEs even for CR7 except for a tentative C {\sc iv} detection (Section \ref{sec_civ}). The flux ratio of the UV-nebular lines relative to Ly$\alpha$ is $f_{\rm UV\,line}/f_{\rm Ly\alpha}\lesssim1-10$\%. 
\end{itemize}

Here we discuss the physical origins of bright LAEs with $\log{L_{\rm Ly\alpha}}/{\rm [erg\ s^{-1}]} \simeq 43-44$. The bright Ly$\alpha$ emission could be reproduced by several mechanisms: (1) gas photo-ionizaiton by a hidden AGN, (2) strong UV radiation from Pop III stellar populations, (3) gas shock heating by strong outflows from central galaxies, and (4) intense starbursts by galaxy mergers. 

Firstly, we discuss the possibility of AGNs. For $z\simeq2$, \citet{2016ApJ...823...20K} have identified a significant hump of LAE number density at the Ly$\alpha$ LF bright-end of $\log{L_{\rm Ly\alpha}}/{\rm [erg\ s^{-1}]} \gtrsim 43.4$. All of the $z\simeq2$ LAEs in the bright-end hump are detected in X-ray, UV, or radio data, suggesting that the bright Ly$\alpha$ emission is produced by the central AGN activity. Similarly, there is a possibility that AGNs enhance the Ly$\alpha$ luminosity for bright LAEs at $z\simeq6-7$. However, we find no clear signatures of AGNs according to the narrow Ly$\alpha$ line widths of $\lesssim400$ km s$^{-1}$ and no detections of N {\sc v} line, X-ray, MIR, nor radio emission. Thus, the bright LAEs at $z\simeq5.7-6.6$ do not host broad-line AGNs, regardless of the bright Ly$\alpha$ emission. 

Secondly, we discuss the possibility of Pop III stellar populations. There is a possibility that strong UV radiation from Pop III stellar populations enhance the Ly$\alpha$ luminosity (e.g., \cite{2002A&A...382...28S}). In our deep NIR spectroscopy, we find that there are no detections of He {\sc ii} emission line from CR7, Himiko, nor our 7 bright LAEs which are observed with NIR spectrographs. Moreover, the Ly$\alpha$ EW does not significantly exceed the $EW_{\rm 0,Ly\alpha}$ value of $240$\,\AA\ for the bright LAEs. The no He {\sc ii} detection and the small $EW_{\rm 0,Ly\alpha}$ values might indicate that the bright LAEs do not host Pop III stellar populations. The no Pop III stellar populations in bright LAEs might be supported by theoretical studies. According to a recent theoretical study of \citet{2017MNRAS.467L..51Y}, Pop III-dominated galaxies at $z\simeq7$ have a Ly$\alpha$ luminosity of $L_{\rm Ly\alpha} \simeq 3.0\times10^{42}-2.1\times10^{43}$ erg s$^{-1}$ which is slightly lower than that of our bright LAEs. However, we cannot obtain the conclusion that Pop III stellar populations exist in bright LAEs from the current data of He {\sc ii} measurements. The detectability of He {\sc ii} emission line would largely depend on the stellar initial mass function of galaxies (see Section \ref{sec_ionizing}). To examine whether bright LAEs host Pop III stellar populations, we require NIR spectra whose depth is $\simeq10\times$ deeper than the current NIR flux limits. 

Thirdly, we discuss the possibility that strong outflows enhance the Ly$\alpha$ luminosity (e.g., \cite{2010MNRAS.408..352D}). If strong outflows exist, expelling high velocity clouds could make Ly$\alpha$ lines broad and Ly$\alpha$ emission spatially extended. Our spectroscopy reveals that bright LAEs have a narrow Ly$\alpha$ emission line of $\Delta V_{\rm FWHM}\lesssim400$ km s$^{-1}$. Our $A_{\rm iso}$ measurements also indicate that most of our bright LAEs show the spatially {\it compact} Ly$\alpha$ emission (see Section \ref{sec_phot} and Table \ref{tab_phys}). The narrow Ly$\alpha$ line width and the spatially compact Ly$\alpha$ emission might suggest no strong gaseous outflow from the bright LAEs. However, we cannot conclude the presence of gaseous outflow based on our current data of optical spectra and NB images due to the resonance nature of Ly$\alpha$ photons. To investigate the presence of gaseous outflow, we have to directly measure velocity shifts of low-ionization metal lines with deep NIR spectra for the rest-frame UV continuum emission (e.g., \cite{2014ApJ...788...74S,2014ApJ...795...33E,2015Natur.523..169E, 2015ApJ...809...89T,2017arXiv170301885S}). 

Finally, we discuss the possibility that intense starbursts driven by galaxy mergers produce the large Ly$\alpha$ luminosity. High spatial resolution imaging observations with {\it Hubble} WFC3 have been conducted for two objects out of the 28 bright LAEs, Himiko and CR7, both of which show multiple subcomponents in the rest-frame UV continuum emission (\cite{2015ApJ...808..139S, 2013ApJ...778..102O}). These multiple subcomponents could be indicative of galaxy mergers (e.g., \cite{2013ApJ...773..153J, 2014ApJ...785...64S, 2016ApJ...819...25K}). However, the galaxy morphology has been unclear for the other 26 bright LAEs in the ground-based and seeing-limited HSC images. 

In summary, the physical origins of bright LAEs have still been unknown. At least we can conclude that the bright Ly$\alpha$ emission is not originated from broad-line AGNs. To obtain the definitive conclusion, we need to systematically perform deep NIR spectroscopy and high spatial resolution imaging observations for a large number of bright LAEs.

\subsection{Relation between UV-nebular Line EW and UV-continuum Luminosity}\label{sec_uvew}

Combining samples of our bright LAEs and faint dropouts at $z\simeq5-7$, we examine the relation between the UV-nebular line EWs of C {\sc iv}, He {\sc ii}, and O {\sc iii}] and UV-continuum luminosity. Figure \ref{fig_muv_ewciv} presents the rest-frame EW of C {\sc iv}, He {\sc ii}, and O {\sc iii}] as a function of $M_{\rm UV}$ for our bright LAEs and dropouts in literature (e.g., \cite{2015MNRAS.454.1393S,2017ApJ...836L..14M, 2017MNRAS.467.3306S}). Here we plot four UV continuum-detected objects out of our 7 bright LAEs whose UV-nebular line EW can be constrained. The EW upper limits of our bright LAEs are typically $\lesssim 2.3$, $4.0$, and $2.9$ \AA\ for C {\sc iv}, He {\sc ii}, and O {\sc iii}] lines, respectively. On the other hand, faint dropouts with $M_{\rm UV}\gtrsim-20$ emit strongly C {\sc iv} and O {\sc iii}] lines with $EW_{\rm 0,CIV}\simeq20-40$\,\AA\ and $EW_{\rm 0,OIII]}\simeq5-10$\,\AA, respectively. 

As shown in Figure \ref{fig_muv_ewciv}, we find a trend that EWs of C {\sc iv} and O {\sc iii}] increase towards faint $M_{\rm UV}$. Such a trend is similar to recent study results for $z\simeq2-3$ galaxies showing that UV-nebular lines are predominantly detected in faint sources (\cite{2014MNRAS.445.3200S, 2017NatAs...1E..52A}; see also \cite{2017ApJ...838...63D} for C {\sc iii}]$\lambda\lambda1907, 1909$). On the other hand, we do not find a clear trend for He {\sc ii} due to no He {\sc ii} detection from all of our bright LAEs nor $z\simeq6-7$ dropouts. For the clarity of the $EW_{\rm 0,CIV}$ and $EW_{\rm 0,OIII]}$ relations, we fit a quadratic function to the  data points of $z\simeq6-7$ dropouts in \citet{2015MNRAS.454.1393S} and \citet{2017ApJ...836L..14M} and our LAEs. In the fitting, we use the values of EW upper limits for the objects without a UV-nebular line detection. We exclude the LAE with a tentative C {\sc iv} detection and a $z\simeq7$ dropout with a weak $EW_{\rm 0,OIII]}$ constraint in \citet{2015MNRAS.454.1393S} for the fitting (see Section \ref{sec_civ}). The best-fit quadratic functions are shown in Figure \ref{fig_muv_ewciv}. 

In contrast to the gravitationally lensed and faint dropouts of \citet{2015MNRAS.454.1393S}, \citet{2017ApJ...836L..14M}, and \citet{2017MNRAS.467.3306S}, our bright LAEs have a moderately bright UV magnitude ranging from $M_{\rm UV}\simeq-20$ to $\simeq-22$. The no UV-nebular line detections from the bright sources could suggest that such a high $EW_0$ value is a characteristic of low-mass galaxies. The high UV-nebular line EW in low-mass galaxies would be due to a hard ionizing spectrum (i.e. $\xi_{\rm ion}$, the number of LyC photons per UV luminosity; e.g., \cite{2016ApJ...831L...9N, 2016ApJ...831..176B}). Moreover, recent studies for $z\simeq0$ galaxies report that high-ionization UV-nebular lines highly depend on the gas-phase metallicity (e.g., \cite{2017arXiv170600881S}). Our possible $EW-M_{\rm UV}$ correlation may also suggest a dependence of UV-nebular line EW on metallicity for $z\simeq6-7$ galaxies via the mass-metallicity relation.

\subsection{A Tentative Detection of C {\sc iv} Emission Line}\label{sec_civ}

In this section, we discuss the EW and UV-nebular line ratios for the LAE whose C {\sc iv} is tentatively detected (Section \ref{sec_fratio}). We estimate the C {\sc iv} EW, $EW_{\rm 0, CIV}$, by using the upper limits of the rest-frame UV continuum flux density. We obtain $EW_{\rm 0, CIV}\gtrsim40$\,\AA, which is comparable to that of a $z\simeq7$ dropout in \citet{2015MNRAS.454.1393S}. The $EW_{\rm 0, CIV}$ value might be too high according to the anti-correlation between $EW$ and UV-continuum luminosity in Section \ref{sec_uvew}. However, it should be noted that the UV continuum is not detected for HSC J233408$+$004403. In the case that the UV magnitude is fainter than $M_{\rm UV}\simeq-21$, the $EW_{\rm 0, CIV}$ value would be comparable to the trend that $EW_0$ is high at a high UV-continuum luminosity. 

\begin{figure}[t!]
 \begin{center}
  \includegraphics[width=80mm]{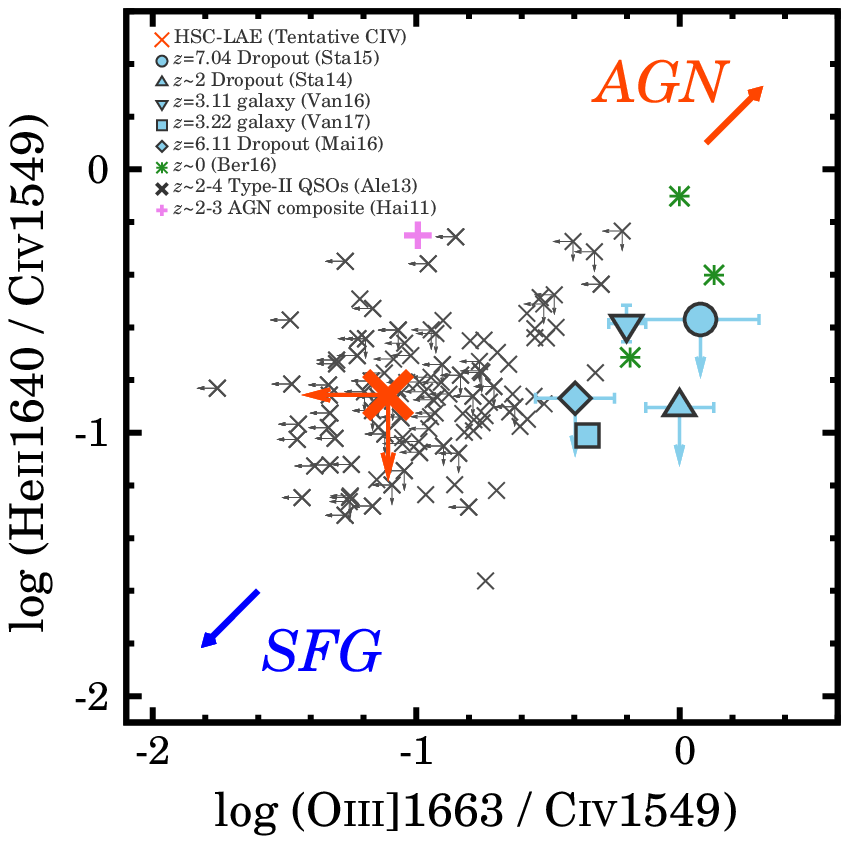}
 \end{center}
 \caption{Flux ratios of UV-nebular emission lines, He {\sc ii}/C {\sc iv} vs O {\sc iii}$]$/C {\sc iv}. The red cross denotes our bright LAE with a tentative C {\sc iv} emission, HSC J233408$+$004403. The cyan filled symbols indicate dropouts at $z\simeq2-7$ (cyan squares: \cite{2017ApJ...842...47V}; cyan filled inverse-triangle: \cite{2016ApJ...821L..27V}; cyan filled diamond; \cite{2017ApJ...836L..14M}; cyan filled triangle: \cite{2014MNRAS.445.3200S}). The green asterisks represent $z\simeq0$ galaxies in \citet{2016ApJ...827..126B}. The crosses present QSOs and AGNs (black crosses: $z\simeq2-4$ type-II QSOs in \cite{2013MNRAS.435.3306A}; magenta cross: $z\simeq2-3$ AGN composite in \cite{2011ApJ...733...31H}). The blue and red arrows indicate the SFG and AGN regions predicted by a photoionization model of \citet{2016MNRAS.456.3354F}, respectively. }\label{fig_line_ratio_uv}
\end{figure}

Assuming that the C {\sc iv} emission line is detected in HSC J233408$+$004403, we compare the He {\sc ii}/C {\sc iv} and O {\sc iii}]/C {\sc iv} line flux ratios of HSC J233408$+$004403 with those of star-forming galaxies at $z\simeq0-7$ and AGNs/QSOs (\cite{2016ApJ...821L..27V,2017ApJ...842...47V, 2017ApJ...836L..14M, 2014MNRAS.445.3200S, 2016ApJ...827..126B, 2013MNRAS.435.3306A, 2011ApJ...733...31H}). Figure \ref{fig_line_ratio_uv} shows the line flux ratios of He {\sc ii}/C {\sc iv} and O {\sc iii}]/C {\sc iv} for HSC J233408$+$004403 and star-forming galaxies and AGNs/QSOs. As shown in Figure \ref{fig_line_ratio_uv}, HSC J233408$+$004403 has a flux ratio limit of log(He {\sc ii}/C {\sc iv})$\lesssim-0.9$ similar to that of star-forming galaxies at $z\simeq7$. 

We compare the limits of flux ratios with those of photoionization models of star-forming galaxies and AGNs in \citet{2016MNRAS.456.3354F}. The comparison suggests that the constraints on the line flux ratios for HSC J233408$+$004403 are more comparable to star-forming galaxies as ionizing sources than AGNs predicted by the model predictions, supporting the results of no clear signatures of AGN activity in Sections \ref{sec_lya_width} and \ref{sec_x_radio}.

\subsection{Spectral Hardness of Bright LAEs}\label{sec_ionizing}

\begin{figure}[t!]
 \begin{center}
  \includegraphics[width=80mm]{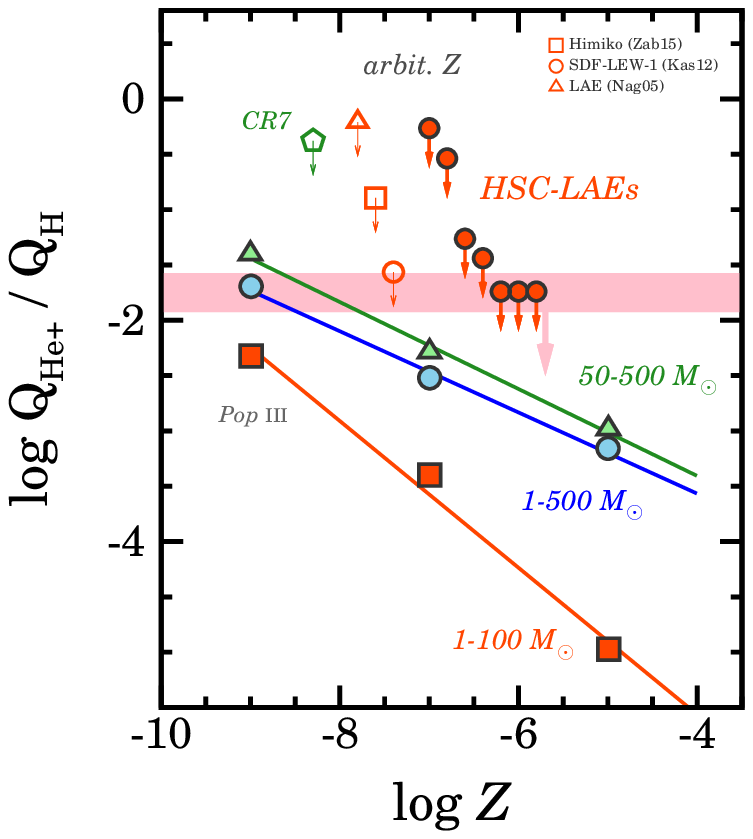}
 \end{center}
 \caption{Spectral hardness of the He$^+$ ionizing flux, $Q_{\rm He^+}/Q_{\rm H}$, as a function of metallicity. The red filled circles represent our bright LAEs. The magenta line indicate the strongest upper limit of our $Q_{\rm He^+}/Q_{\rm H}$ estimates. The filled red squares, blue circles, and green triangles with colored lines denote the model predictions of \citet{2003A&A...397..527S} for stellar initial mass functions with mass ranges of $1-100$ M$_\odot$, $1-500$ M$_\odot$, and $50-500$ M$_\odot$, respectively. The open symbols indicate $z\gtrsim6$ LAEs (red open square: Himiko in \cite{2015MNRAS.451.2050Z}; red open circle: SDF-LEW-1 in \cite{2012ApJ...761...85K}; SDF J132440.6$+$273607: \cite{2005ApJ...631L...5N}). The open green pentagon is $Q_{\rm He^+}/Q_{\rm H}$ obtained from our He {\sc ii}/Ly$\alpha$ constraint for CR7 (see Section \ref{sec_cr7}). The metallicity of the observational data points is arbitrary. The data points without a UV-nebular line detection indicate $2\sigma$ upper limits.}\label{fig_metal_hardness}
\end{figure}

We investigate the spectral hardness of bright LAEs at $z\simeq6-7$ based on the upper limits on the He {\sc ii}/Ly$\alpha$ line flux ratios (Section \ref{sec_fratio}). Figure \ref{fig_metal_hardness} presents the spectral hardness, $Q_{\rm He+}/Q_{\rm H}$, as a function of metallicity, $Z$, for our bright LAEs and $z\simeq6-7$ LAEs in previous studies (Himiko in \cite{2015MNRAS.451.2050Z}; SDF-LEW-1 in \cite{2012ApJ...761...85K}; SDF J132440.6$+$273607 in \cite{2005ApJ...631L...5N}). Here we use $Q_{\rm He+}/Q_{\rm H}$ which is more model-independent than physical quantities of e.g., $\xi_{\rm ion}$. The $Q_{\rm He+}/Q_{\rm H}$ value is calculated with an equation of  

\begin{equation}\label{eq_hardness}
 \frac{f_{\rm He}}{f_{\rm Ly\alpha}} \simeq 0.55 \times \frac{Q({\rm He}^+)}{Q({\rm H})}, 
\end{equation}

\noindent where, $f_{\rm He}$ and $f_{\rm Ly\alpha}$ are the flux of He {\sc ii} and Ly$\alpha$ emission line, respectively. $Q({\rm He}^+)$ and $Q({\rm H})$ are the emitted number of hydrogen and helium ionizing photons, respectively. The $Q_{\rm He+}/Q_{\rm H}$ traces the energy range between 54.4 and 13.6 eV. The factor of $0.55$ depends on the electron temperature, here taken to be $T_{\rm e} = 30$ kK (\cite{2002A&A...382...28S}). The $Q ({\rm He}^+)/Q({\rm H})$ upper limits calculated from the He {\sc ii}/Ly$\alpha$ line flux ratios (Table \ref{tab_uvline}) ranges from $\log{Q({\rm He}^+)/Q({\rm H})}\simeq-0.5$ to $\simeq-1.8$. For five objects of our bright LAEs, we put strong upper limits of $\log{Q ({\rm He}^+)/Q({\rm H})}\lesssim-1.8$. 

Figure \ref{fig_metal_hardness} also shows the the model spectral hardness predicted from initial mass functions (IMFs) with different stellar mass ranges of $M_*=1-100$ M$_\odot$, $1-500$ M$_\odot$, and $M_*=50-500$ M$_\odot$(\cite{2003A&A...397..527S}). The metallicity of bright LAEs has not been constrained yet. If we assume that bright LAEs are extremely metal poor below $\log{Z}\simeq-8$, top-heavy IMFs with $M_*=50-500$ M$_\odot$ might be ruled out by our $Q_{\rm He+}/Q_{\rm H}$ constraints for $z\simeq6-7$ LAEs.

\section{Summary and Conclusions}\label{sec_summary}

We present Ly$\alpha$ and UV-nebular emission line properties of bright LAEs at $z=6-7$ with a luminosity of $\log{L_{\rm Ly\alpha}/{\rm [erg\ s^{-1}]}} =43-44$ identified in the 21-deg$^2$ area of the SILVERRUSH early sample developed with the Subaru/HSC survey data (\cite{2017arXiv170407455O, 2017arXiv170408140S}). 

Our findings are summarized as follows: 

\begin{itemize} 
\item Our optical spectroscopy newly confirm 21 bright LAEs with clear Ly$\alpha$ emission, and contribute to make a spectroscopic sample of 97 LAEs at $z=6-7$ in SILVERRUSH. Our observations enlarge a spectroscopic sample of bright LAEs by a factor of four, allowing for a statistical study on bright LAEs. We find that all the bright LAEs have a narrow Ly$\alpha$ line width of $\lesssim400$ km s$^{-1}$, and do not have X-ray, MIR, radio, nor N {\sc v}$\lambda\lambda 1238, 1240$ emissions regardless of the large Ly$\alpha$ luminosity. The narrow Ly$\alpha$ line widths and no X-ray, MIR, radio, nor N {\sc v} detections suggest that the bright LAEs are not broad-line AGNs. 

\item From the spectroscopic sample, we select 7 remarkable LAEs as bright as Himiko and CR7 objects, and perform deep Keck/MOSFIRE and Subaru/nuMOIRCS NIR spectroscopy reaching the $3\sigma$-flux limit of $\simeq 2\times 10^{-18}$ erg s$^{-1}$ for the UV-nebular emission lines of He {\sc ii}$\lambda1640$, C {\sc iv}$\lambda\lambda1548,1550$, and O {\sc iii}]$\lambda\lambda1661,1666$. Except for one tentative detection of C {\sc iv}, we find no strong UV-nebular lines down to the flux limit, placing the upper limits of $EW_0$ of $\sim 2.3$, $4.0$, and $2.9$ \AA\ for He {\sc ii}, C {\sc iv}, and O {\sc iii}] lines, respectively. 

\item We investigate the VLT/X-SHOOTER spectrum of CR7 whose $6 \sigma$ detection of He {\sc ii} is claimed by \citet{2015ApJ...808..139S}. Although two individuals of the authors in this paper and the ESO-archive service carefully re-analyze the X-SHOOTER data that are used in the study of \citet{2015ApJ...808..139S}, no He {\sc ii} signal of CR7 is detected, supportive of weak UV-nebular lines of the bright LAEs even for CR7. 

\item Spectral properties of these bright LAEs are clearly different from those of faint dropouts at $z\sim 7$ that have strong UV-nebular lines shown in the various studies (e.g., \cite{2015MNRAS.454.1393S}). Comparing these bright LAEs and the faint dropouts, we find anti-correlations between the UV-nebular line $EW_0$ and UV-continuum luminosity, which are similar to those found at $z\sim 2-3$.

\end{itemize} 

The high spatial resolution imaging and deep spectroscopic observations with {\it Hubble Space Telescope} and {\it James Webb Space Telescope} will reveal the morphology, ISM properties, and the origins of bright LAEs.

\section{Appendix}\label{sec_appendix}

Tables \ref{tab_faint_z66} and \ref{tab_faint_z57} present faint ${\it NB}>24$ spectroscopically confirmed HSC LAEs at $z\simeq6.6$ and $z\simeq5.7$, respectively. See Section \ref{sec_confirm} for more details.

\begin{longtable}{*{7}{c}}
\caption{Spectroscopically confirmed $z\simeq6.6$ LAEs with $NB>24$ mag}\label{tab_faint_z66}
\hline
Object ID & $\alpha(J2000)$ & $\delta(J2000)$ & $z_{\rm spec}$ & ${\it NB}921$ & {\it y} & Reference \\
 & & & & (mag) & (mag) &  \\
(1) & (2) & (3) & (4) & (5) & (6) & (7) \\
\hline
\endfirsthead
\endhead
\hline
\endfoot
\hline 
\multicolumn{7}{l}{(1) Object ID.} \\
\multicolumn{7}{l}{(2) Right ascension.} \\
\multicolumn{7}{l}{(3) Declination. } \\
\multicolumn{7}{l}{(4) Spectroscopic redshift of Ly$\alpha$ emission line. } \\
\multicolumn{7}{l}{(5)-(6) Total magnitudes of ${\it NB}921$ and $y$-bands. } \\
\multicolumn{7}{l}{(7) Reference ({\tt O10}: \cite{2010ApJ...723..869O}; {\tt Hari}: Y. Harikane in prep.; {\tt H}: R. Higuchi in prep.). } \\
\multicolumn{7}{l}{Note that the magnitudes are values directly obtained from the HSC catalog. } \\
\endlastfoot
\multicolumn{7}{c}{${\it NB}921$ ($z\simeq6.6$)} \\ \hline 
HSC J021843$-$050915 & 02:18:43.62 & $-$05:09:15.63 & 6.510 & 24.33 & 24.87 & Hari \\ 
HSC J021703$-$045619 & 02:17:03.46 & $-$04:56:19.07 & 6.589 & 24.45 & 25.42 & O10 \\ 
HSC J021827$-$043507 & 02:18:27.01 & $-$04:35:07.92 & 6.511 & 24.56 & 25.32 & O10 \\ 
HSC J021844$-$043636 & 02:18:44.64 & $-$04:36:36.21 & 6.621 & 24.63 & 27.34 & H \\ 
HSC J021702$-$050604 & 02:17:02.56 & $-$05:06:04.61 & 6.545 & 24.64 & 26.35 & O10 \\ 
HSC J021826$-$050726 & 02:18:27.00 & $-$05:07:26.89 & 6.554 & 24.69 & --- & O10 \\ 
HSC J021819$-$050900 & 02:18:19.39 & $-$05:09:00.65 & 6.563 & 24.73 & 26.04 & O10 \\ 
HSC J021654$-$045556 & 02:16:54.54 & $-$04:55:56.94 & 6.617 & 24.82 & 25.67 & O10 \\ 
\hline 
\end{longtable}

\begin{longtable}{*{7}{c}}
\caption{Spectroscopically confirmed $z\simeq5.7$ LAEs with $NB>24$ mag}\label{tab_faint_z57}
\hline
Object ID & $\alpha(J2000)$ & $\delta(J2000)$ & $z_{\rm spec}$ & ${\it NB}816$ & {\it z} & Reference \\
 & & & & (mag) & (mag) &  \\
(1) & (2) & (3) & (4) & (5) & (6) & (7) \\
\hline
\endfirsthead
\endhead
\hline
\endfoot
\hline 
\multicolumn{7}{l}{(1) Object ID.} \\
\multicolumn{7}{l}{(2) Right ascension.} \\
\multicolumn{7}{l}{(3) Declination. } \\
\multicolumn{7}{l}{(4) Spectroscopic redshift of Ly$\alpha$ emission line. } \\
\multicolumn{7}{l}{(5)-(6) Total magnitudes of ${\it NB}816$ and $z$-bands. } \\
\multicolumn{7}{l}{(7) Reference ({\tt O08}: \cite{2008ApJS..176..301O}; {\tt H}: R. Higuchi in prep.). } \\
\multicolumn{7}{l}{Note that the magnitudes are values directly obtained from the HSC catalog. } \\
\endlastfoot
\multicolumn{7}{c}{${\it NB}816$ ($z\simeq5.7$)} \\ \hline 
HSC J095952$+$013723 & 09:59:52.13 & $+$01:37:23.24 & 5.724 & 24.07 & 25.76 & M12 \\ 
HSC J021758$-$043030 & 02:17:58.91 & $-$04:30:30.42 & 5.689 & 24.07 & 25.56 & H \\ 
HSC J095933$+$024955 & 09:59:33.44 & $+$02:49:55.92 & 5.724 & 24.10 & 27.25 & M12 \\ 
HSC J021749$-$052854 & 02:17:49.11 & $-$05:28:54.17 & 5.694 & 24.10 & 26.77 & O08 \\ 
HSC J021704$-$052714 & 02:17:04.30 & $-$05:27:14.30 & 5.686 & 24.11 & 26.29 & H \\ 
HSC J095952$+$015005 & 09:59:52.03 & $+$01:50:05.95 & 5.744 & 24.11 & 25.10 & M12 \\ 
HSC J021737$-$043943 & 02:17:37.96 & $-$04:39:43.02 & 5.7547 & 24.11 & 25.63 & H \\ 
HSC J100015$+$020056 & 10:00:15.66 & $+$02:00:56.04 & 5.718 & 24.15 & 26.08 & M12 \\ 
HSC J021734$-$044558 & 02:17:34.57 & $-$04:45:58.95 & 5.702 & 24.20 & 25.44 & H \\ 
HSC J100131$+$023105 & 10:01:31.08 & $+$02:31:05.77 & 5.690 & 24.23 & 26.15 & M12 \\ 
HSC J100301$+$020236 & 10:03:01.15 & $+$02:02:36.04 & 5.682 & 24.24 & 24.58 & M12 \\ 
HSC J021654$-$052155 & 02:16:54.60 & $-$05:21:55.52 & 5.712 & 24.24 & 26.49 & H \\ 
HSC J021748$-$053127 & 02:17:48.46 & $-$05:31:27.02 & 5.690 & 24.25 & 25.67 & O08 \\ 
HSC J100127$+$023005 & 10:01:27.77 & $+$02:30:05.83 & 5.696 & 24.28 & 25.61 & M12 \\ 
HSC J021745$-$052936 & 02:17:45.24 & $-$05:29:36.01 & 5.688 & 24.30 & 27.26 & O08 \\ 
HSC J021725$-$050737 & 02:17:25.90 & $-$05:07:37.59 & 5.704 & 24.35 & 26.21 & H \\ 
HSC J100208$+$015444 & 10:02:08.80 & $+$01:54:44.99 & 5.676 & 24.36 & 25.65 & M12 \\ 
HSC J095954$+$021039 & 09:59:54.77 & $+$02:10:39.26 & 5.662 & 24.38 & 25.63 & M12 \\ 
HSC J095950$+$025406 & 09:59:50.09 & $+$02:54:06.16 & 5.726 & 24.39 & 26.59 & M12 \\ 
HSC J022013$-$045109 & 02:20:13.33 & $-$04:51:09.40 & 5.744 & 24.40 & 25.88 & O08 \\ 
HSC J100126$+$014430 & 10:01:26.88 & $+$01:44:30.29 & 5.686 & 24.41 & 25.96 & M12 \\ 
HSC J095919$+$020322 & 09:59:19.74 & $+$02:03:22.02 & 5.704 & 24.41 & 26.84 & M12 \\ 
HSC J095954$+$021516 & 09:59:54.52 & $+$02:15:16.50 & 5.688 & 24.43 & 25.95 & M12 \\ 
HSC J021849$-$052235 & 02:18:49.00 & $-$05:22:35.35 & 5.719 & 24.45 & 25.64 & H \\ 
HSC J100005$+$020717 & 10:00:05.06 & $+$02:07:17.01 & 5.704 & 24.46 & 26.64 & M12 \\ 
HSC J021830$-$052950 & 02:18:30.75 & $-$05:29:50.34 & 5.707 & 24.46 & 28.89 & H \\ 
HSC J100306$+$014742 & 10:03:06.13 & $+$01:47:42.69 & 5.680 & 24.52 & 26.54 & M12 \\ 
HSC J021804$-$052147 & 02:18:04.17 & $-$05:21:47.25 & 5.7338 & 24.54 & 25.20 & H \\ 
HSC J100022$+$024103 & 10:00:22.51 & $+$02:41:03.25 & 5.661 & 24.55 & 25.34 & M12 \\ 
HSC J021848$-$051715 & 02:18:48.23 & $-$05:17:15.45 & 5.741 & 24.56 & 25.45 & H \\ 
HSC J021750$-$050203 & 02:17:50.86 & $-$05:02:03.24 & 5.708 & 24.57 & 26.48 & H \\ 
HSC J021526$-$045229 & 02:15:26.22 & $-$04:52:29.93 & 5.655 & 24.62 & 24.95 & H \\ 
HSC J021636$-$044723 & 02:16:36.44 & $-$04:47:23.68 & 5.718 & 24.63 & 26.57 & H \\ 
HSC J100030$+$021714 & 10:00:30.41 & $+$02:17:14.73 & 5.695 & 24.65 & 26.70 & M12 \\ 
HSC J021558$-$045301 & 02:15:58.49 & $-$04:53:01.75 & 5.718 & 24.68 & 26.55 & H \\ 
HSC J021719$-$043150 & 02:17:19.13 & $-$04:31:50.64 & 5.735 & 24.68 & 27.87 & H \\ 
HSC J021822$-$042925 & 02:18:22.91 & $-$04:29:25.89 & 5.697 & 24.68 & 27.65 & H \\ 
HSC J100131$+$014320 & 10:01:31.11 & $+$01:43:20.50 & 5.728 & 24.70 & 26.45 & M12 \\ 
HSC J095944$+$020050 & 09:59:44.07 & $+$02:00:50.74 & 5.688 & 24.71 & 26.18 & M12 \\ 
HSC J021709$-$050329 & 02:17:09.77 & $-$05:03:29.18 & 5.709 & 24.74 & 26.52 & H \\ 
HSC J021803$-$052643 & 02:18:03.87 & $-$05:26:43.45 & 5.747 & 24.75 & 27.66 & H \\ 
HSC J100309$+$015352 & 10:03:09.81 & $+$01:53:52.36 & 5.705 & 24.76 & 26.61 & M12 \\ 
HSC J021805$-$052704 & 02:18:05.17 & $-$05:27:04.06 & 5.746 & 24.77 & 31.43 & H \\ 
HSC J021739$-$043837 & 02:17:39.25 & $-$04:38:37.21 & 5.720 & 24.79 & 27.00 & H \\ 
HSC J100040$+$021903 & 10:00:40.24 & $+$02:19:03.70 & 5.719 & 24.81 & 26.96 & M12 \\ 
HSC J021857$-$045648 & 02:18:57.32 & $-$04:56:48.88 & 5.681 & 24.85 & 27.11 & H \\ 
HSC J021745$-$044129 & 02:17:45.74 & $-$04:41:29.24 & 5.674 & 24.86 & 27.34 & H \\ 
HSC J021639$-$051346 & 02:16:39.89 & $-$05:13:46.75 & 5.702 & 24.87 & 26.98 & H \\ 
HSC J021805$-$052026 & 02:18:05.28 & $-$05:20:26.90 & 5.742 & 24.87 & 26.10 & H \\ 
HSC J021755$-$043251 & 02:17:55.40 & $-$04:32:51.54 & 5.691 & 24.91 & 27.26 & H \\ 
HSC J100058$+$013642 & 10:00:58.41 & $+$01:36:42.89 & 5.688 & 24.91 & 27.97 & M12 \\ 
HSC J100029$+$015000 & 10:00:29.58 & $+$01:50:00.78 & 5.707 & 24.97 & 26.80 & M12 \\ 
HSC J021911$-$045707 & 02:19:11.03 & $-$04:57:07.48 & 5.704 & 25.00 & 27.46 & H \\ 
HSC J021551$-$045325 & 02:15:51.34 & $-$04:53:25.44 & 5.710 & 25.02 & 26.76 & H \\ 
HSC J021625$-$045237 & 02:16:25.64 & $-$04:52:37.18 & 5.728 & 25.07 & --- & H \\ 
HSC J021751$-$053003 & 02:17:51.14 & $-$05:30:03.64 & 5.712 & 25.10 & 26.99 & O08 \\ 
HSC J021628$-$050103 & 02:16:28.05 & $-$05:01:03.85 & 5.692 & 25.17 & 27.23 & H \\ 
HSC J021943$-$044914 & 02:19:43.91 & $-$04:49:14.30 & 5.684 & 25.17 & 26.86 & H \\ 
HSC J100029$+$024115 & 10:00:29.13 & $+$02:41:15.70 & 5.735 & 25.22 & 28.30 & M12 \\ 
HSC J100107$+$015222 & 10:01:07.35 & $+$01:52:22.88 & 5.668 & 25.33 & 26.42 & M12 \\ 
\hline 
\end{longtable}

\begin{ack}
 We would like to thank Masayuki Akiyama, Mark Dijkstra, Richard Ellis, Tadayuki Kodama, Jorryt Matthee, David Sobral, Daniel Stark, Yuma Sugahara, and Zheng Zheng for useful discussion and comments. We also thank Kentaro Aoki and Ichi Tanaka for their supports of the FOCAS and MOIRCS observations. We thank the anonymous referee for constructive comments and suggestions. This work is based on observations taken by the Subaru Telescope and the Keck telescope which are operated by the National Observatory of Japan. This work was supported by World Premier International Research Center Initiative (WPI Initiative), MEXT, Japan, KAKENHI (15H02064), (23244025), and (21244013) Grant-in-Aid for Scientific Research (A) through Japan Society for the Promotion of Science (JSPS), and an Advanced Leading Graduate Course for Photon Science grant. NK is supported by JSPS grant 15H03645. 
 
 The Hyper Suprime-Cam (HSC) collaboration includes the astronomical communities of Japan and Taiwan, and Princeton University. The HSC instrumentation and software were developed by the National Astronomical Observatory of Japan (NAOJ), the Kavli Institute for the Physics and Mathematics of the Universe (Kavli IPMU), the University of Tokyo, the High Energy Accelerator Research Organization (KEK), the Academia Sinica Institute for Astronomy and Astrophysics in Taiwan (ASIAA), and Princeton University. Funding was contributed by the FIRST program from Japanese Cabinet Office, the Ministry of Education, Culture, Sports, Science and Technology (MEXT), the Japan Society for the Promotion of Science (JSPS), Japan Science and Technology Agency (JST), the Toray Science Foundation, NAOJ, Kavli IPMU, KEK, ASIAA, and Princeton University. 

This paper makes use of software developed for the Large Synoptic Survey Telescope. We thank the LSST Project for making their code available as free software at  http://dm.lsst.org

The Pan-STARRS1 Surveys (PS1) have been made possible through contributions of the Institute for Astronomy, the University of Hawaii, the Pan-STARRS Project Office, the Max-Planck Society and its participating institutes, the Max Planck Institute for Astronomy, Heidelberg and the Max Planck Institute for Extraterrestrial Physics, Garching, The Johns Hopkins University, Durham University, the University of Edinburgh, Queen's University Belfast, the Harvard-Smithsonian Center for Astrophysics, the Las Cumbres Observatory Global Telescope Network Incorporated, the National Central University of Taiwan, the Space Telescope Science Institute, the National Aeronautics and Space Administration under Grant No. NNX08AR22G issued through the Planetary Science Division of the NASA Science Mission Directorate, the National Science Foundation under Grant No. AST-1238877, the University of Maryland, and Eotvos Lorand University (ELTE) and the Los Alamos National Laboratory.
\end{ack}

\bibliographystyle{apj}
\bibliography{reference_prop}

\end{document}